\PassOptionsToPackage{pdfpagelabels=false, citecolor=true}{hyperref}
\documentclass[12pt,apj,preprint,iop]{emulateapj}
\usepackage{longtable,array}
\newlength\mylength
\newcolumntype{C}[1]{>{\centering\arraybackslash}p{#1}}
\usepackage{natbib}
\usepackage{amsmath}
\usepackage{hyperref}
\hypersetup{
    colorlinks=true,
    citecolor=blue,
    linkcolor=magenta,
    urlcolor=cyan
}

\def\rmxaa{\ref@jnl{Rev. Mexicana Astron. Astrofis.}}

\usepackage[normalem]{ulem}
\usepackage{color}
\usepackage{setspace}

\shorttitle{PNe Color Identification in Broadband Surveys}
\shortauthors{Vejar et al. 2018}

\begin{document}

\title{Planetary Nebulae and How to Find Them: Color Identification in Big Broadband Surveys}

\author{George Vejar\altaffilmark{1,5}, Rodolfo Montez Jr.\altaffilmark{2}, Margaret Morris\altaffilmark{4} and Keivan G. Stassun\altaffilmark{1,3}}
\affil{\altaffilmark{1}Department of Physics and Astronomy, Vanderbilt University, Nashville, TN 37235}
\affil{\altaffilmark{2} Center for Astrophysics $\vert$ Harvard \& Smithsonian, Cambridge, MA 02138, USA}
\affil{\altaffilmark{3}Department of Physics, Fisk University, Nashville, TN
37208, USA}
\affil{\altaffilmark{4}Scripps Institution of Oceanography, UC San Diego, San Diego, CA 92037, USA}
\affil{\altaffilmark{5}LSSTC Data Science Fellow}

\email{george.vejar@vanderbilt.edu}

\begin{abstract}
Planetary nebulae (PNe) provide tests of stellar evolution, can serve as tracers of chemical evolution in the Milky Way and other galaxies, and are also used as a calibrator of the cosmological distance ladder.
Current and upcoming large scale photometric surveys have the potential to 
complete the census of PNe in our galaxy and beyond, but it is a challenge to disambiguate partially or fully unresolved PNe from the myriad other sources observed in these surveys.
Here we carry out synthetic observations of nebular models to determine $ugrizy$ color-color spaces that can successfully identify PNe among billions of other sources. 
As a primary result we present a grid of synthetic absolute magnitudes for PNe at various stages of their evolution, and we make comparisons with real PNe colors from the Sloan Digital Sky Survey.
We find that the $r-i$ versus $g-r$, and the $r-i$ versus $u-g$, color-color diagrams show the greatest promise for cleanly separating PNe from stars, background galaxies, and quasars.
Finally, we consider the potential harvest of PNe from upcoming large surveys.
For example, for typical progenitor host star masses of $\sim$3~M$_\odot$, we find that the Large Synoptic Survey Telescope (LSST) should be sensitive to virtually all PNe in the Magellanic Clouds with extinction up to $A_{\rm V}$ of $\sim$5~mag; out to the distance of Andromeda, LSST would be sensitive to the youngest PNe (age less than $\sim$6800~yr) and with $A_{\rm V}$ up to 1 mag. 
\end{abstract}


\section{Introduction}

Planetary Nebulae (PNe) are the shells of gas ionized by hot central stars.
The PN forms from previously ejected material lost during the asymptotic giant branch (AGB) phase of a low-to-intermediate mass star ($\rm1-8~M_{\odot}$). 
The central star of a planetary nebula (CSPN) plays a key role in the PN characteristics since its fast stellar wind plows into the AGB wind to form the nebula while the CSPN's high surface temperature photoionizes the gas in the newly-formed nebula. 
A PN will expand and fade over time, while the CSPN rises in temperature, until it eventually cools towards the white dwarf (WD) cooling track \citep{1978ApJ...219L.125K,1994ApJS...92..125V,1995AA...299..755B, 2016AA...588A..25M}.
Compared to the lifetime of the star, the PN phase is short-lived remaining visible for only $\sim10^{4}$ years \citep{1995PhR...250....1I,2013AA...558A..78J}.

For over 60 years the formation process of PNe has been questioned favoring two contending processes.
The first consists of a single star in which heavy mass loss occurs during the AGB phase while stellar rotation and magnetic fields shape the expanding nebula \citep{1962VA......5...40G,1999ApJ...517..767G, 2005ApJ...618..919G, 2004ASPC..313..449M, 2004ASPC..313..401B}.
The second process favors a wide range of interactions between the evolving star and a binary companion for shaping the nebula \citep{1979MNRAS.187..283F, 1997ApJS..112..487S, 2009PASP..121..316D}.
Theoretical and observational considerations suggest that a single AGB star is unlikely to produce a strong enough magnetic field to dramatically shape the nebula, favouring the possibility of binary interactions being mainly responsible for the formation and shaping of non-spherical PN \citep{2006PASP..118..260S,2007MNRAS.376..599N}.
It is likely that there are still many PNe left undiscovered as our best estimates place the total number of galactic PNe anywhere between $\sim 6600$ and $4.6\pm1.3 \times10^{4}$ depending on the formation process \citep{2005AIPC..804..169D,2006ApJ...650..916M}. 

Naturally, because they represent a specific and short-lived phase of stellar evolution, PNe can be difficult to study but they are important for improving our understanding of late-stage low- to intermediate-mass stars \citep{1995PhR...250....1I,2008PhDT.......109F}.
PNe that have been found in the Milky Way and in neighboring galaxies have been extremely valuable for a variety of studies.
Because of their bright emission lines, PNe are identifiable across the Galaxy and in nearby stellar systems.
PNe can be used as tracers of galaxy kinematics \citep{2004ApJ...616..804H,2004ApJ...602..705P,2006MNRAS.369..120M,2016IAUFM..29B..20C}; as a rung on the distance ladder via the PN luminosity function \citep{2003astro.ph..1279C,2018NatAs...2..580G}; and as potential tracers of the chemical evolution of the Milky Way and other galaxies \citep{1997ApJ...487..651W,1998AA...340...67R,2008MNRAS.388.1667K,2009AA...494..515S,2009MNRAS.398..280M,2012ApJ...753...12K,2012ApJ...758..133S,2017MNRAS.468..272C}.

There is ongoing interest in techniques for readily identifying more PNe efficiently and reliably.
A large number of PNe have been discovered through techniques that take advantage of their bright nebular emission lines.
$\rm H\alpha$ surveys such as the SuperCOSMOS $\rm H\alpha$ Survey (SHS) \citep{2005MNRAS.362..689P,2014MNRAS.440.1080F}, the INT Photometric $\rm H\alpha$ Survey (IPHAS) \citep{2005MNRAS.362..753D}, and the VST Photometric $\rm H\alpha$ Survey (VPHAS+) \citep{2014MNRAS.440.2036D} have been very successful in identifying galactic PNe and are cataloged in the Hong Kong/AAO/Strasbourg/$\rm H\alpha$ (HASH) \citep{2016JPhCS.728c2008P} database which contains $\sim 3500$ objects.
Integral field spectroscopy of [\ion{O}{3}] $\lambda5007$ has been used to find PNe in crowded areas \citep{2013MNRAS.430.1219P}.
Dust can make it difficult for optical surveys to detect PNe but they have also been found in the UKIRT Wide-field Imaging Survey for $\rm H_{2}$ survey (UWISH2) through their $\rm H_{2}$ emission \citep{2018MNRAS.479.3759G}.
More galactic PNe have recently been discovered through their multi-wavelength characteristics ranging from optical to radio emission \citep{2018MNRAS.tmp.1882F}.

Large, all-sky surveys, like the Sloan Digital Sky Survey (SDSS) and, in the future, the Large Synoptic Survey Telescope (LSST), can be used to potentially identify thousands of new PN based on broadband photometry where PNe are poorly characterized.
Although the observed broadband colors of PNe have been studied \citep[e.g., in 2MASS and WISE colors][]{2001AA...377L..18S,2010PASA...27..129F,2017AA...606A.110I}, and although the theoretically expected broadband colors of PN central stars have previously been calculated \citep[e.g.,][]{2009JPhCS.172a2033W,2015AAS...22513907M}, the theoretically expected broadband colors of PN nebular emission have yet to be characterized and compared with observations.
Existing PNe catalogs are far from complete, and upcoming all-sky photometric surveys have great potential to uncover troves of additional PNe, assuming that it will be possible to efficiently distinguish true PNe from the large numbers of false positives.

We focus on the optical region of the PN spectrum because of the prominent nebular emission lines at these wavelengths.
A vast amount of $ugriz(y)$ survey data is or will become available through SDSS and LSST, therefore we attempt to characterize PNe in this photometric system. 
We consider the broadband $ugrizy$ characteristics of a synthetic PN as a function of evolutionary age, for both resolved and unresolved PNe.
Our methodology for creating our synthetic PN models and calculating the synthesized $ugrizy$ observations are presented in \S\ref{sec:methods}. 
A grid of broadband absolute magnitudes and the efficacy of different color-color diagrams to reliably identify PNe are provided in \S\ref{sec:results}. 
In \S\ref{sec:discussion} we compare our results to existing observational SDSS studies and consider future applications with LSST. 
Finally, \S\ref{sec:summary} presents a summary of our conclusions.

\section{Methodology}\label{sec:methods}

In this section, we describe our methodology for creating synthetic broadband $ugrizy$ observations of PNe as functions of evolutionary age for both resolved and unresolved PNe. 
We begin by describing the physical ingredients and assumptions adopted. 
Because we seek to undertake a more comprehensive study of the broadband behavior of PNe than previously attempted, we intentionally construct our methodology to fully and self-consistently include the effects of both the central star and the nebular evolution.
Next we describe the production of the nebular spectra followed by the synthetic $ugrizy$ magnitudes and colors from these spectra, and we discuss our treatment of resolved and unresolved PNe.
Finally, we include the effects of interstellar reddening for resolved and unresolved spectra.

\subsection{Planetary Nebula Ingredients}

\begin{figure}[!ht]
\centering
\includegraphics[width=3.4 in]{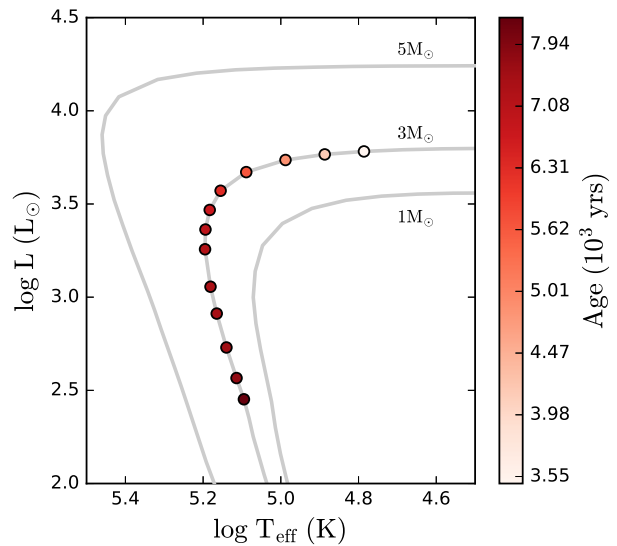}
\caption[Central Star Evolutionary Tracks]{Hertzsprung-Russell diagram showing the evolution of the central star for various planetary nebula progenitor masses (gray lines). The points represent luminosity and temperature parameters used for our 13 Cloudy models for a 3$M_{\odot}$ progenitor star. The shading of the color represents the age of the model, refer to Table \ref{table: PNe Parameters} for parameter values.}
\label{fig:HR}
\end{figure}

\begin{figure}[!ht]
\centering
\includegraphics[width=3.4 in]{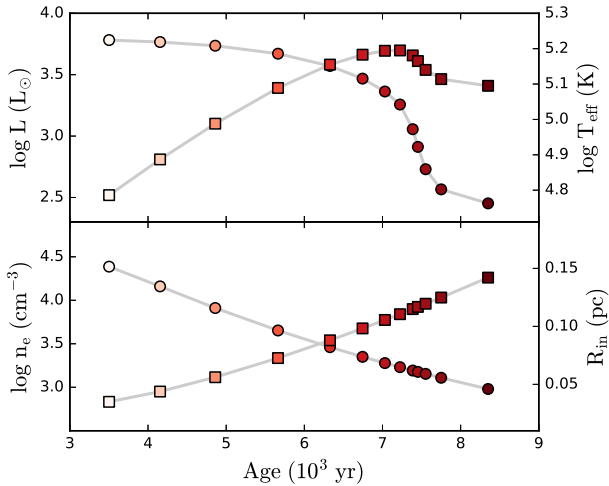}
\caption[Evolution of Nebula Properties]{(Top panel): Evolution of central star over time; squares correspond to temperature and circles correspond to luminosity. (Bottom panel): Evolution of the nebula over time; squares correspond to the inner radius of the nebula and circles correspond to the electron density. The shading of the color represents the age of the model, refer to Table \ref{table: PNe Parameters} for parameter values.}
\label{fig:nebula_evolution}
\end{figure}

To build our synthetic PN we start with the central star and nebular properties.
In particular, throughout this study we work with a 3~$M_{\odot}$ progenitor star, because it is for such a star that nebular evolution models exist to which we can self-consistently couple the evolution of the central star and thereby produce synthetic spectra and colors over the course of the evolution of the star+PN.
Figure~\ref{fig:HR} depicts the HR diagram for a $3\ M_{\odot}$ central star evolutionary track from the models of \citet{1994ApJS...92..125V,1995AA...299..755B}, along with $1\ M_{\odot}$ and $5\ M_{\odot}$ progenitor masses for context.
The final mass of the white dwarf in this model is ${0.605~M_{\odot}}$, similar to the $\approx$0.6~M$_\odot$ mass that is often assumed as a fiducial mass for PN central stars \citep{2004AA...414..993P}.
Thirteen specific age positions on the evolutionary track are indicated, representing the discrete ages that we have chosen to include in our study, spanning the evolution of the system from emergence of the PN to exposing the white dwarf. 
Figure~\ref{fig:nebula_evolution} (top panel) indicates the temporal behavior of the CSPN luminosity ($L$) and temperature ($T_{\rm eff}$) corresponding to these 13 ages.

For every age position on the central star's evolutionary track (Figure~\ref{fig:HR}) we 
require a nebular emission model corresponding to that evolutionary age and with the appropriate central star mass.
Fortunately, 1D radiation-hydrodynamics simulations of a PN have been performed by \citet{2004AA...414..993P,2005AA...441..573S}. Because those simulations were done for the expansion of a PN illuminated by a central remnant of 0.595~$M_{\odot}$, it is essential to use a central-star evolutionary model that produces a remnant as close to this same final mass as possible. That is why our choice above is of a central star whose final mass is most similar to this value\footnote{The slight discrepancy between the central star final masses of 0.595~$M_{\odot}$ versus 0.605~$M_{\odot}$ in the nebular versus central star evolutionary tracks, respectively, is negligible for compact nebula with radii $< 1 {\rm ~pc}$ \citep{2013AA...558A..78J}.}, corresponding to a progenitor mass of 3~$M_\odot$.

We calculate the nebular size at each time step using a simple approximation based on the \citet{2004AA...414..993P,2005AA...441..573S} nebular model. 
Based on the radiation-hydrodynamic simulations, the densest part of the nebula is coincident with the inner radius (or rim), $R_{\rm in}$, which follows the relationship $R = R_{0}+c(t-t_{0})^{b}$, where $R_0 = 0.031$~pc, $t_0 = 3016$~yrs, $c = 6.7 \times 10^{-7}{\rm ~pc~yr}^{-1}$, and $b = 1.4$. 
For each time step in the evolutionary track, we determine the inner radius ($R_{\rm in}$) based on this prescription.
Since the prescription is only applicable for $3016<t<10000$~yrs, we only considered the evolutionary track between these ages. 
As a result, the radii of our synthetic nebula model expands from $0.035$~pc to $0.142$~pc.

The outer radius $R_{\rm out}$ of the nebula is determined by Cloudy calculations (see next section), specifically by a mass stopping criteria for each model.
\citet{2007MNRAS.378..231P} describe an empirical relationship to the total ionized mass, $M_{\rm ionized}$. 
$M_{\rm ionized}$ increases linearly for $R \leq 0.1 {\rm ~pc}$ according to $M_{\rm ionized} = M_0 (R/0.1{\rm ~pc})$ until reaching a constant total mass, $M_{0}$, for $R>0.1{\rm ~pc}$. 
In our radiative transfer calculations, we restrict the total ionized mass of the nebula according to this empirical relationship using $M_{0}\approx0.2\ M_{\odot}$. 
As a result the final mass of our synthetic nebula are designed to closely match the final masses of the ionized nebula described in \citet{2007MNRAS.378..231P}.
Also as a consequence of this stopping mass criteria the thickness of the nebular shell remains thin, specifically, $(R_{\rm out}-R_{\rm in})/R_{\rm in}$ is always $<$ 0.25.

In addition to the inner radius, for each considered position on the evolutionary track we must calculate the  electron density, $n_{e}$, of the nebula. 
Based on the behavior of $\sim240$ Galactic and Magellanic PNe \citet{2008PhDT.......109F} derived a nebular density relation with nebular radius of $\log n_{e} = -2.31(\pm0.04)\log R+1.02(\pm0.04)$, where $n_{e}$ is the electron density of the nebula and $R$ is the nebular radius, in our case $R_{\rm in}$.
Figure~\ref{fig:nebula_evolution} (bottom panel) shows the evolution of $n_{e}$ and $R_{\rm in}$ over all ages of the nebula.

\subsection{Cloudy, With a Chance of Photons}\label{sec:cloudycalcs}

The central star and nebular properties described in the previous section are used as input to Cloudy radiative transfer calculations.
Cloudy is a plasma simulation software that simulates non-equilibrium gas conditions to predict an observable spectrum(version 13.03; \cite{2013RMxAA..49..137F}).

\begin{figure*}
\centering
\includegraphics[width=6 in]{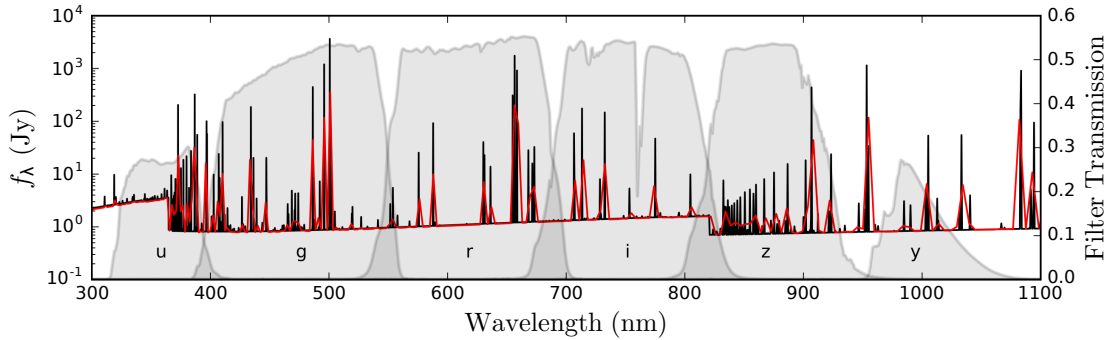}
\caption[Coarse Cloudy Output Spectrum]{Coarse (standard) resolution spectrum of the youngest model age = $3502$~yrs as output by Cloudy ($T_{\star}=61,094$~K, ${L_{\star}}=6,053~L_{\odot}$) (red) along with the high resolution spectrum for the same model (black). The coarse spectrum is used for calculating the magnitudes in each filter. Overlaid in gray is a representative $ugrizy$ LSST filter set (described in the text). Note that the $y$ filter is not used in this study.}
\label{fig:coarse_spectrum}
\end{figure*}

In Table~\ref{table: PNe Parameters}, we list the stellar and nebular ingredients needed to run our Cloudy radiative transfer models. 
For the nebular composition we use chemical abundances typical of a planetary nebula \citep[see Table~\ref{table: PN abunds}][]{2013RMxAA..49..137F,1983ApJS...51..211A,1989SSRv...51..339K}. 
We assume spherical geometry for the nebula, a uniform filling factor of unity, and no stopping temperature criteria. 
Distance between the observer and the nebula is assumed to be 1 kpc. 
The resulting coarse (R = 200) and high-resolution (R = 2000) spectra at the youngest nebular age considered ($3502$~yrs, $61094$~K, $6053~L_{\odot}$) are shown in Figure~\ref{fig:coarse_spectrum} for our region of interest ($300 {\rm ~to~} 1100{\rm ~nm}$).
The high-resolution spectrum is used for display purposes.
Since the high-resolution spectrum does not preserve flux, we use the coarse spectrum for our photometric calculations. 

\begin{table}[!ht]
\centering
\begin{tabular}{c c c c c c}
    \hline
    $Age$ & $\log~T_{\star}$ & $\log~\frac{L_{\star}}{L_{\odot}}$ & $R_{in}$ & $\log~n_{e}$ & $M_{stop}$ \\
    (yrs) & (K) & ($L_{\odot}$) & (pc) & ($cm^{-3}$) & ($M_{\odot}$) \\
    \hline
    3502  & 4.786 & 3.782 & 0.035 & 4.386 & 0.071 \\
    4154  & 4.887 & 3.766 & 0.044 & 4.159 & 0.088 \\
    4860  & 4.988 & 3.736 & 0.056 & 3.911 & 0.113 \\
    5663  & 5.089 & 3.671 & 0.073 & 3.652 & 0.146 \\
    6328  & 5.155 & 3.571 & 0.088 & 3.460 & 0.178 \\
    6745  & 5.183 & 3.468 & 0.098 & 3.349 & 0.199 \\
    7031  & 5.194 & 3.363 & 0.106 & 3.277 & 0.201 \\
    7224  & 5.195 & 3.257 & 0.111 & 3.230 & 0.201 \\
    7387  & 5.181 & 3.056 & 0.115 & 3.192 & 0.202 \\
    7453  & 5.165 & 2.912 & 0.117 & 3.176 & 0.202 \\
    7552  & 5.140 & 2.73  & 0.119 & 3.153 & 0.203 \\
    7751  & 5.114 & 2.566 & 0.125 & 3.108 & 0.202 \\
    8351  & 5.095 & 2.452 & 0.142 & 2.980 & 0.202 \\
    \hline
\end{tabular}
\caption[PN Parameters]{Parameters used for the 13 Cloudy models that change from model to model. These parameters represent the evolution of the central star ($T_{\star}$, $L_{\star}$) and nebula ($R_{in}$, $n_{e}$) while $M_{stop}$ is used to stop the simulation once the nebula has accumulated enough mass.}
\label{table: PNe Parameters}
\end{table}

\begin{table}[!ht]
\centering
\begin{tabular}{c c}
    \hline
	Atom & [X/H] \\
	 & (dex) \\
	\hline
	H  &  0.0000 \\
	He & -1.0000 \\
	C  & -3.1079 \\
	N  & -3.7447 \\
	O  & -3.3565 \\
	F  & -6.5229 \\ 
	Ne & -3.9586 \\
	Na & -5.7212 \\
	Mg & -5.7959 \\
	Al & -6.5686 \\
	Si & -5.0000 \\
	P  & -6.6990 \\  
	S  & -5.0000 \\  
	Cl & -6.7696 \\  
	Ar & -5.5686 \\  
	K  & -6.9208 \\  
	Ca & -7.9208 \\  
	Fe & -6.3010 \\
	Ni & -7.7447 \\
    \hline
\end{tabular}
\caption[PN Abundances]{Abundances used for PN models within Cloudy \citep{1983ApJS...51..211A,1989SSRv...51..339K}. Values are relative to solar values from \citet{1998SSRv...85..161G},
\citet{2001AIPC..598...23H}, \citet{2001ApJ...556L..63A,2002ApJ...573L.137A}. Elements not mentioned are assumed to be depleted enough to be of no consequence. For more information on available abundances within Cloudy refer to \textit{``Hazy"} the Cloudy manual \citep{2013RMxAA..49..137F}.}
\label{table: PN abunds}
\end{table}

\subsection{Synthesizing Broadband Photometric Observations}
\hspace*{3mm}

We consider observations of our synthetic nebulae performed with a standard $ugrizy$ photometric system\footnote{We used the LSST filters as defined on 2016-12-07 obtained from the Spanish Virtual Observatory (SVO) \href{http://svo2.cab.inta-csic.es/theory/fps3/index.php?mode=browse}{Filter Profile Service}. The $y$ filter is shown for completeness but is not used in this study.}. 
The set of photometric filters cover wavelengths between 320 nm and 1080 nm and their properties are provided in Table~\ref{table: Filter Properties} with their transmission curves shown in  Figure~\ref{fig:coarse_spectrum}.
The filter transmission curve is interpolated onto the wavelength grid of the coarse resolution Cloudy spectra described in \S{\ref{sec:cloudycalcs}}. 
For each age of our synthetic nebula we convolved the nebular and central star flux spectra with the filter transmission curves. 
Then we normalized the resulting flux to the AB magnitude system to determine the $ugrizy$ magnitudes. 
We scaled these magnitudes to 10 pc to determine the absolute magnitudes, which are presented in Table~\ref{table: PNe Magnitudes}. 
\begin{table}[!ht]
\centering
\begin{tabular}{c c c c c c}
    \hline
    Filter & Range & $\lambda_{\rm eff}$ & $W_{\rm eff}$ & $m_{sat}$ & $m_{5\sigma}$\\
     & (nm) & (nm) & (nm) & (mags) & (mags) \\
    \hline
    $u$ & 320--400 & 373.2 & 54.6  & 14.7 & 23.9  \\
    $g$ & 400--552 & 473.0 & 133.2 & 15.7 & 25.0 \\
    $r$ & 552--691 & 613.8 & 133.7 & 15.8 & 24.7 \\
    $i$ & 691--818 & 748.7 & 832.5 & 15.8 & 24.0 \\
    $z$ & 818--922 & 866.8 & 937.5 & 15.3 & 23.3 \\
    $y$\tablenotemark{a} & 950--1080 & 967.6 & 81.0 & 13.9 & 22.1 \\
    \hline
\end{tabular}
\caption[ugrizy Filter Properties]{Properties of the $ugrizy$ filter set used in our analysis. All magnitudes are normalized to the AB magnitude system \citep{1983ApJ...266..713O}.}
\tablenotetext{0}{$^a$ Not used in this study.}
\label{table: Filter Properties}
\end{table}

\subsection{Accounting for Spatially Resolved Extended Emission}

The extended nature of a PN requires additional consideration. 
For aperature photometry, when the nebula and central star are resolved only a fraction of the nebular flux will be measured. 
To understand the impact of resolved nebula and central stars, we constructed a toy model of the synthetic nebulae. 
For each stage, we assumed a three-dimensional spherical shell with the radius, thickness, and density used in our synthetic nebulae. 
We projected this shell onto the plane of the sky. 
Assuming aperture photometry, we considered a range of aperture diameters on the sky, $\Omega_{\rm aperture}$, and nebular diameter on the sky, $\Omega_{\rm neb}$. 
For a range of aperture to nebular diameter ratios,  $\Omega_{\rm aperture}/\Omega_{\rm neb}$, we calculated the fraction of nebular emission, $f_{\rm neb}$, that is measured by the aperture. 
We recalculated the magnitudes after scaling the nebular flux by this fraction and adding it to the central star flux.
We consider models with the same nebular fraction as a fractional variant and label them with roman numerals in increasing order as $f_{\rm neb}$ decreases. 

This method of estimating resolved nebular magnitudes provides the most flexibility with regards to survey resolution characteristics.
Note that because in reality a given PN, at a given distance, will evolve in its apparent angular size, it will not in general evolve along a single variant ``track". Rather, the collection of variant tracks represent the overall parameter space through which individual PNe may evolve, for a range of distances, photometric apertures, and states of evolution.

\subsection{Reddening} 

\begin{figure*}
\centering
\includegraphics[width=6 in]{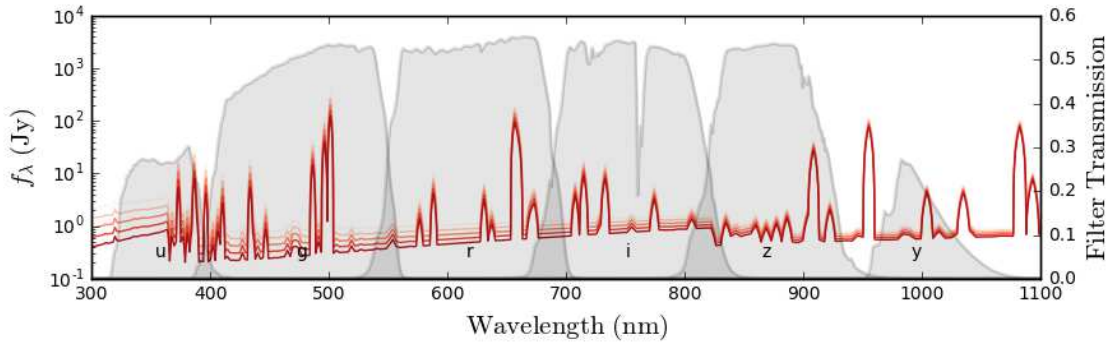}
\caption[Spectrum Reddening]{Youngest model age = $3502$~yrs ($T_{\star}=61,094$~K, ${L_{\star}}=6,053~L_{\odot}$) spectrum shown with varying levels of extinction ($\rm A_{V}$) along with the LSST filter set. Lines are shaded lightest to darkest for $\rm A_{V} = 0,~0.25,~0.5,~0.75,~1.0$.}
\label{fig:extinct_spectrum}
\end{figure*}

Intervening dust poses a serious problem for observing and identifying objects with photometric colors. 
The dust will dim (extinction) and redden an object which can limit the distance at which an object can be detected and change its position in a color-color diagram.
To determine the reddening vector in the ugrizy system, we applied the reddening curve of \citet{1989ApJ...345..245C} to all ages of the synthetic nebular spectra then calculated the reddened magnitudes. 
Figure~\ref{fig:extinct_spectrum} shows how varying levels of extinction ($\rm A_{\rm V}$) can affect the spectrum of a PN.
Figures~\ref{fig:cmd} and \ref{fig:colorcolor} show the reddening vector that all 13 models would follow with $A_{\rm V}=1 {\rm ~mag}$.

\section{Results}\label{sec:results}

In this section, we present the main results of our study.
First, we provide a grid of synthetic $ugrizy$ photometry representing our model PN at different evolutionary ages and different observed angular sizes.
Second, we describe the evolution of prominent emission lines in each filter and their effect on the calculated magnitudes. 
We then identify the $ugrizy$ color-color spaces that are most effective at differentiating PNe from other celestial objects.

\subsection{Synthesized Nebula Absolute Magnitudes}

\begin{figure}[!ht]
\centering
\includegraphics[width=3 in]{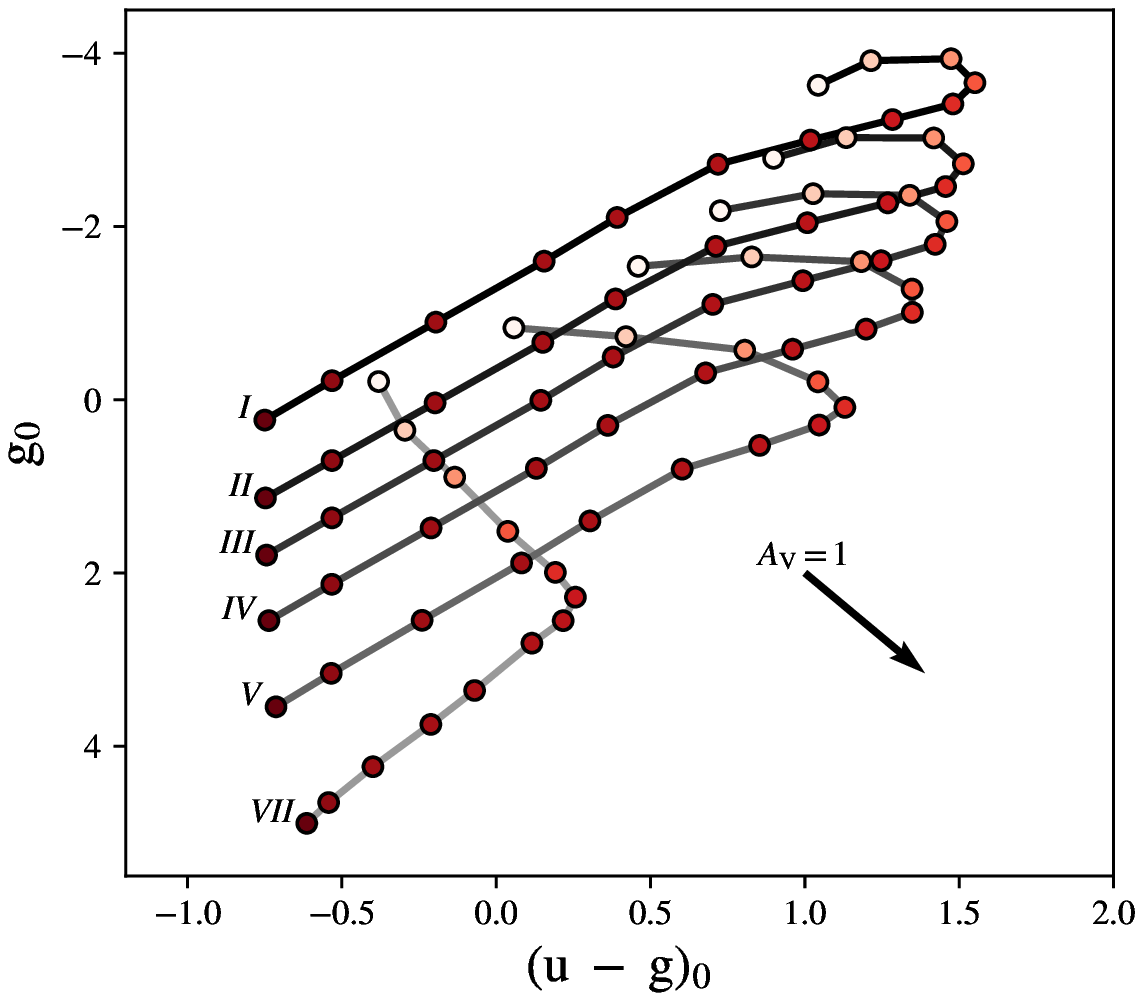}
\includegraphics[width=3 in]{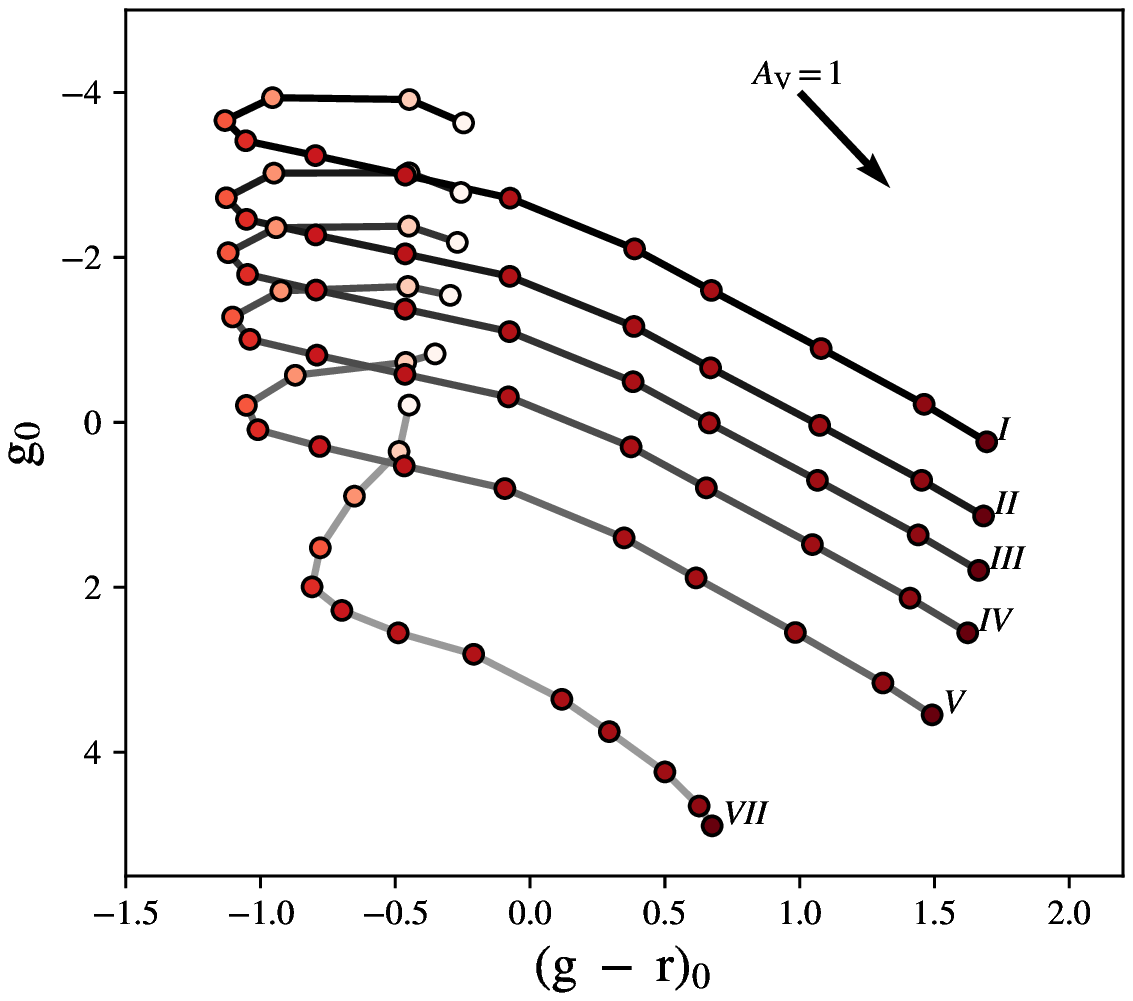}
\includegraphics[width=3 in]{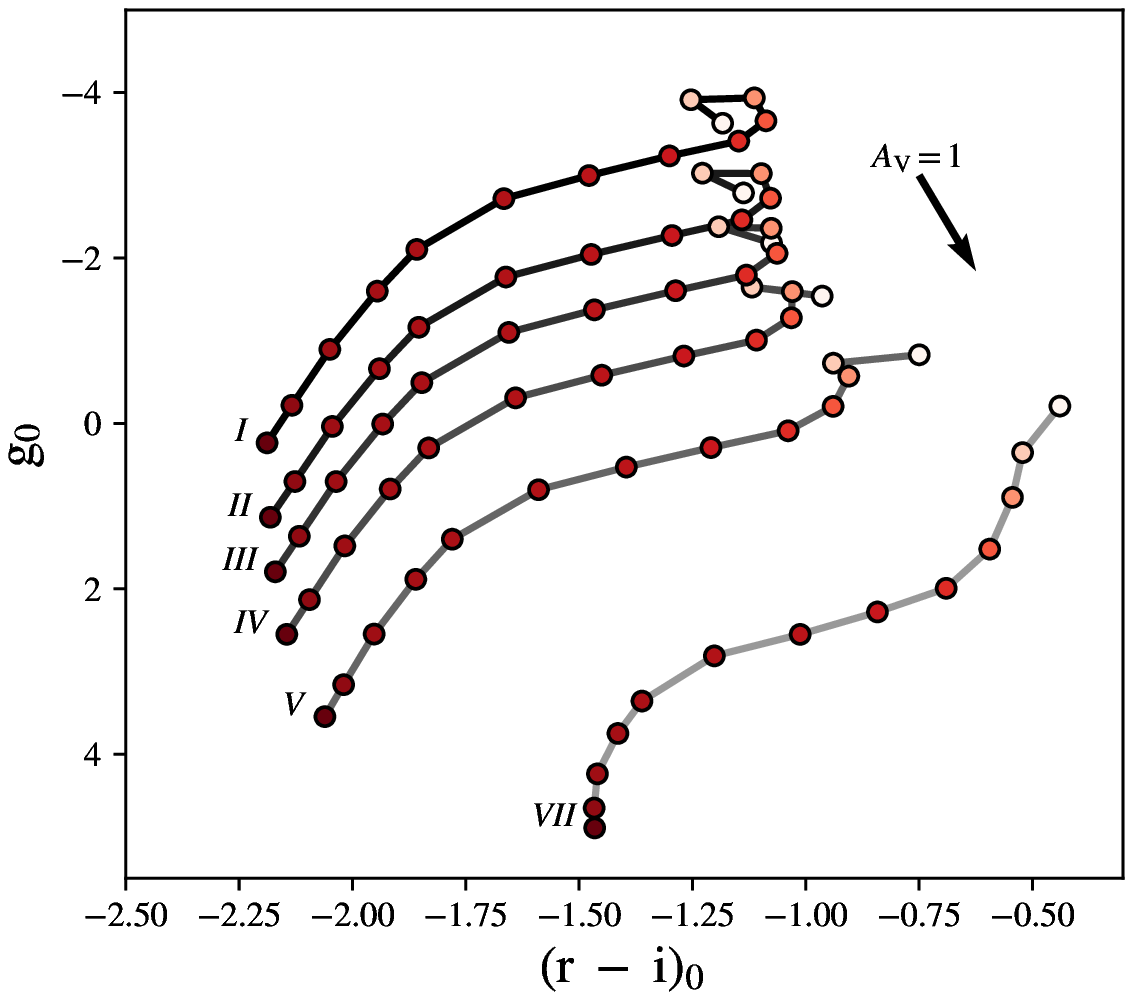}
\caption[Color-Magnitude Diagrams]{Selected color-magnitude diagrams for PN models.
Points are colored similarly to Figure~\ref{fig:HR}.
Grey-scaled shaded lines connect models of the same $f_{\rm neb}$ as indicated by the Roman numeral variant labels.
A black line connects unresolved models, the shade lightens as $f_{\rm neb}$ decreases (see Table~\ref{table: neb_fraction} for values of each variant).}
\label{fig:cmd}
\end{figure}

A key result of this study is the collection of the absolute magnitudes for our synthetic nebulae in the $ugrizy$ photometric system (see Table~\ref{table: PNe Magnitudes}, Table~\ref{table: neb_fraction} and Figure \ref{fig:cmd} for all variants). 
These absolute magnitudes span $\sim4800$ years of PN evolution for a $\rm 3~M_{\odot}$ progenitor star evolving into a $\rm 0.605~M_{\odot}$ white dwarf.
The contributions from continuum and line emission dictate the behavior of the broad band magnitudes as a function of age. 
In each band, the continuum drops by about an order of magnitude as the models approach older stages resulting in the overall decline in the broadband magnitudes as the nebula evolves. 
We do not display variants beyond VII, as variants VIII--X become too faint to be practical.

The line emission as a function of age in each band is often dictated by one or two species that dominate a given bandpass. 
In the $u$ band, the prominent emission lines are: [\ion{Ne}{3}] ($\lambda3869$, $\lambda3968$), [\ion{O}{2}] ($\lambda3726$, $\lambda3729$), \ion{H}{2} ($\lambda3835$, $\lambda3889$, $\lambda3970$), and \ion{He}{3} ($\lambda3203$).
In the $g$ band, the prominent emission lines are: [\ion{O}{3}] ($\lambda5007$, $\lambda4959$), \ion{H}{1} ($\rm H_{\rm \beta}$ $\lambda4861$, $\rm H_{\rm \gamma}$ $\lambda4340$, $\rm H_{\rm \delta}$ $\lambda4102$), \ion{He}{2} ($\lambda4686$), \ion{He}{1} ($\lambda4471$), [\ion{S}{2}] ($\lambda4074$, $\lambda4070$, $\lambda4078$), and \ion{N}{1} ($\lambda5200$, $\lambda5198$).
In the $r$ band, the prominent emission lines are: \ion{H}{1} ($\rm H_{\rm \alpha}$ $\lambda6563$), [\ion{N}{2}] ($\lambda6584$, $\lambda6548$), \ion{He}{1} ($\lambda5876$), [\ion{S}{2}] ($\lambda6720$, $\lambda6731$, $\lambda6716$),  \ion{S}{3} ($\lambda6312$), [\ion{Cl}{3}] ($\lambda5538$) and [\ion{O}{1}] ($\lambda6300$).
In the $i$ band, the prominent emission lines are: [\ion{Ar}{3}] ($\lambda7135$, $\lambda7751$), [\ion{O}{2}] ($\lambda7323$, $\lambda7332$), \ion{He}{1} ($\lambda7065$), and \ion{Cl}{4} ($\lambda8047$).
Fluxes from the aforementioned lines exhibit three common behaviors: (1) {\it decreasing} throughout the evolution, (2) {\it double-peaked} with an initial rise, a decrease, then another rise as the CSPN $T_{\rm eff}$ approaches its hottest temperature before decreasing again as $T_{\rm eff}$ cools, and, (3) {\it single-peak} with a steady increase in flux as the as CSPN $T_{\rm eff}$ approaches its hottest temperature before decreasing as $T_{\rm eff}$ cools.
Lines with decreasing behavior are: [\ion{Ne}{3}], \ion{H}{1}, [\ion{O}{3}], \ion{He}{1}, \ion{S}{3}, [\ion{Cl}{3}], \ion{Cl}{4}, [\ion{Ar}{3}].
Lines with a double peak behavior are: [\ion{O}{2}], [\ion{S}{2}], \ion{N}{1}, [\ion{N}{2}], [\ion{O}{1}].
Only \ion{He}{2} lines show a single peak behavior. 
These behaviors of the line emission are due to the ionization and physical parameters (radius and density) of the modeled nebula. 

When convolved with the bandpass transmission curves the broadband magnitudes can mimic the age-related behavior of the strongest emission lines. 
The $u$ band magnitude behavior is double-peaked with a steady decrease largely influenced by the [\ion{O}{2}] emission and the overall decreasing flux from the other lines in this bandpass. 
The [\ion{Ne}{3}] lines are very close to the edge of the $u$ filter bandpass, where the transmission is reduced, therefore, despite being the most prominent line in the $u$ band, [\ion{Ne}{3}] does not contribute as much flux as the [\ion{O}{2}] lines.
The $g$ band magnitude exhibits decreasing behavior and is largely influenced by [\ion{O}{3}].
The $r$ band magnitude is double-peaked, which is the result of the \ion{H}{1} emission dominating at earlier ages ($t_{\rm age}<7000~{\rm yr}$) until [\ion{N}{2}] begins to dominate. 
The $i$ band magnitude decreases according to the evolution of [\ion{Ar}{3}] with some influence from [\ion{O}{2}].

In the CMDs (Figure \ref{fig:cmd}) there exists a hook feature before the eighth model ($t_{\rm age} \leq 7224$ yrs) (also somewhat visible in the CCDs) due to the evolution of these lines.
Once the models have reached the empirically observed maximum mass we imposed (at the eighth model $t_{\rm age} = 7224$ yrs) they become matter bounded and transparent to ionizing radiation resulting in a somewhat steady decline in brightness and color evolution. 

The absolute magnitudes given in Table~\ref{table: PNe Magnitudes} are only valid for the total integrated flux from a PN. 
When a PN is resolved, only a fraction of the nebula will be observed, thus reducing the nebular contribution. 
This effect changes the measured flux in each filter and the measured color for any two filters.
Figures \ref{fig:cmd}, \ref{fig:colorcolor}, and \ref{fig:Yuan pne} show how magnitudes and colors change as the aperture encloses less of the nebula.

\begin{table}[!ht]
\centering
\begin{tabular}{c c c c c c c}
    \multicolumn{2}{c}{\text{Variant I:}} & \multicolumn{3}{c}{\text{$\Omega_{\rm aperture}/\Omega_{\rm neb} = 1.3$}} & \multicolumn{2}{c}{\text{$f_{neb}$ = $1$}} \\
    \hline
    Age & $u$ & $g$ & $r$ & $i$ & $z$ & $y$\\
    (yrs)  &   (mags) &   (mags) &   (mags) &   (mags) &   (mags) &   (mags) \\
    \hline
    3502 & -2.59 & -3.63 & -3.38 & -2.20 & -2.18 & -2.09 \\
    4154 & -2.70 & -3.91 & -3.47 & -2.21 & -2.20 & -2.15 \\
    4860 & -2.46 & -3.94 & -2.98 & -1.87 & -1.98 & -1.94 \\
    5663 & -2.11 & -3.66 & -2.53 & -1.44 & -1.62 & -1.66 \\
    6328 & -1.93 & -3.41 & -2.36 & -1.21 & -1.40 & -1.49 \\
    6745 & -1.95 & -3.23 & -2.43 & -1.14 & -1.29 & -1.39 \\
    7031 & -1.98 & -3.00 & -2.53 & -1.06 & -1.13 & -1.28 \\
    7224 & -2.00 & -2.72 & -2.64 & -0.98 & -0.98 & -1.21 \\
    7387 & -1.71 & -2.10 & -2.49 & -0.63 & -0.54 & -0.95 \\
    7453 & -1.44 & -1.60 & -2.27 & -0.33 & -0.18 & -0.72 \\
    7552 & -1.09 & -0.90 & -1.98 &  0.08 &  0.28 & -0.45 \\
    7751 & -0.75 & -0.22 & -1.68 &  0.45 &  0.71 & -0.24 \\
    8351 & -0.51 &  0.24 & -1.46 &  0.73 &  1.00 & -0.11 \\
    \hline
\end{tabular}
\caption[PNe Synthesized Absolute Magnitudes]{Predicted absolute magnitudes of our synthetic nebular spectra for the $ugrizy$ filter set. No extinction is applied. Models are fully unresolved with $\Omega_{\rm aperture}/\Omega_{\rm neb} = $ 1.3 and $f_{neb} = $ 1. For varying $\Omega_{\rm aperture}/\Omega_{\rm neb}$ and $f_{neb}$ values see Table \ref{table: neb_fraction}.}
\label{table: PNe Magnitudes}
\end{table}

\subsection{Identifying PNe in $ugrizy$ Color-Color Diagrams}

\begin{figure}[!ht]
\centering
\includegraphics[width=3in]{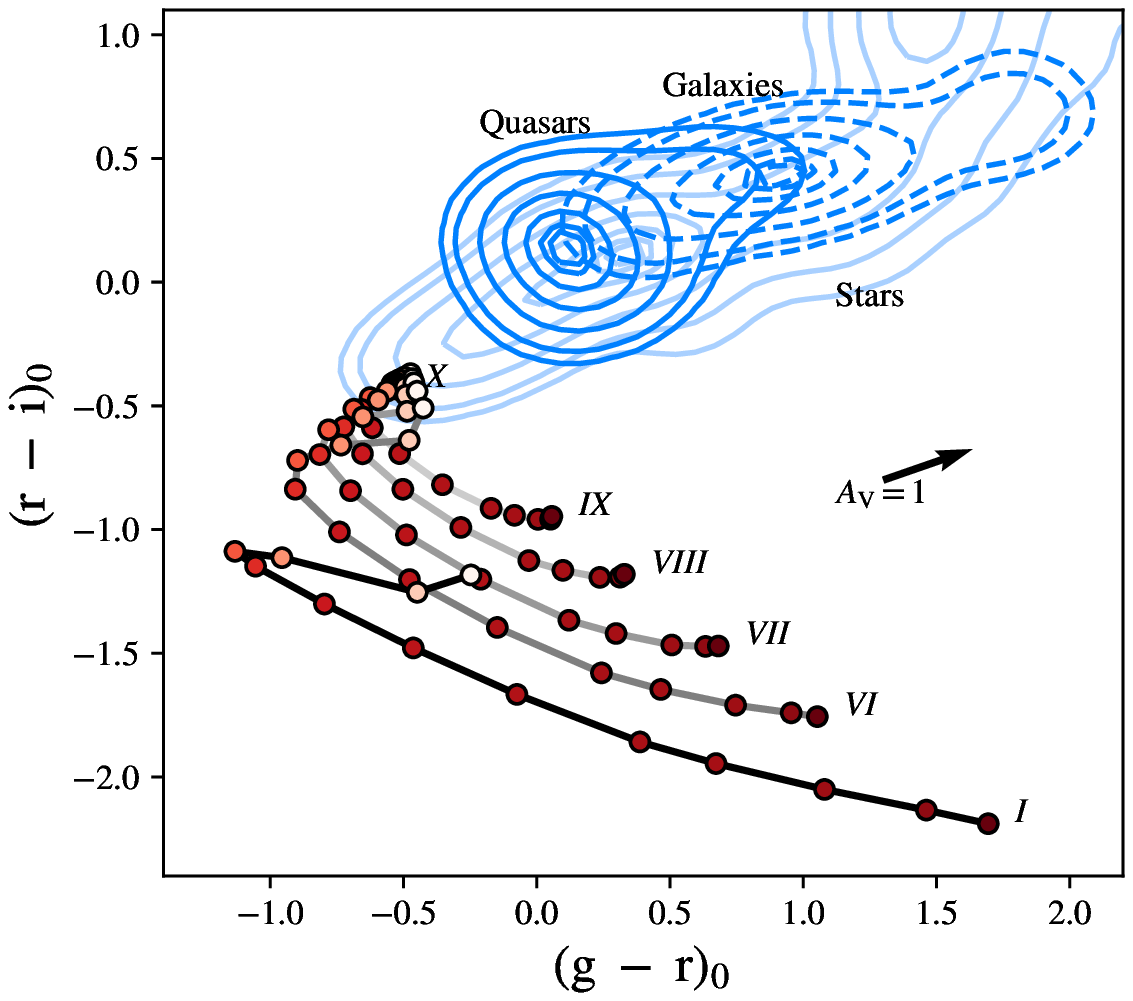}
\includegraphics[width=3in]{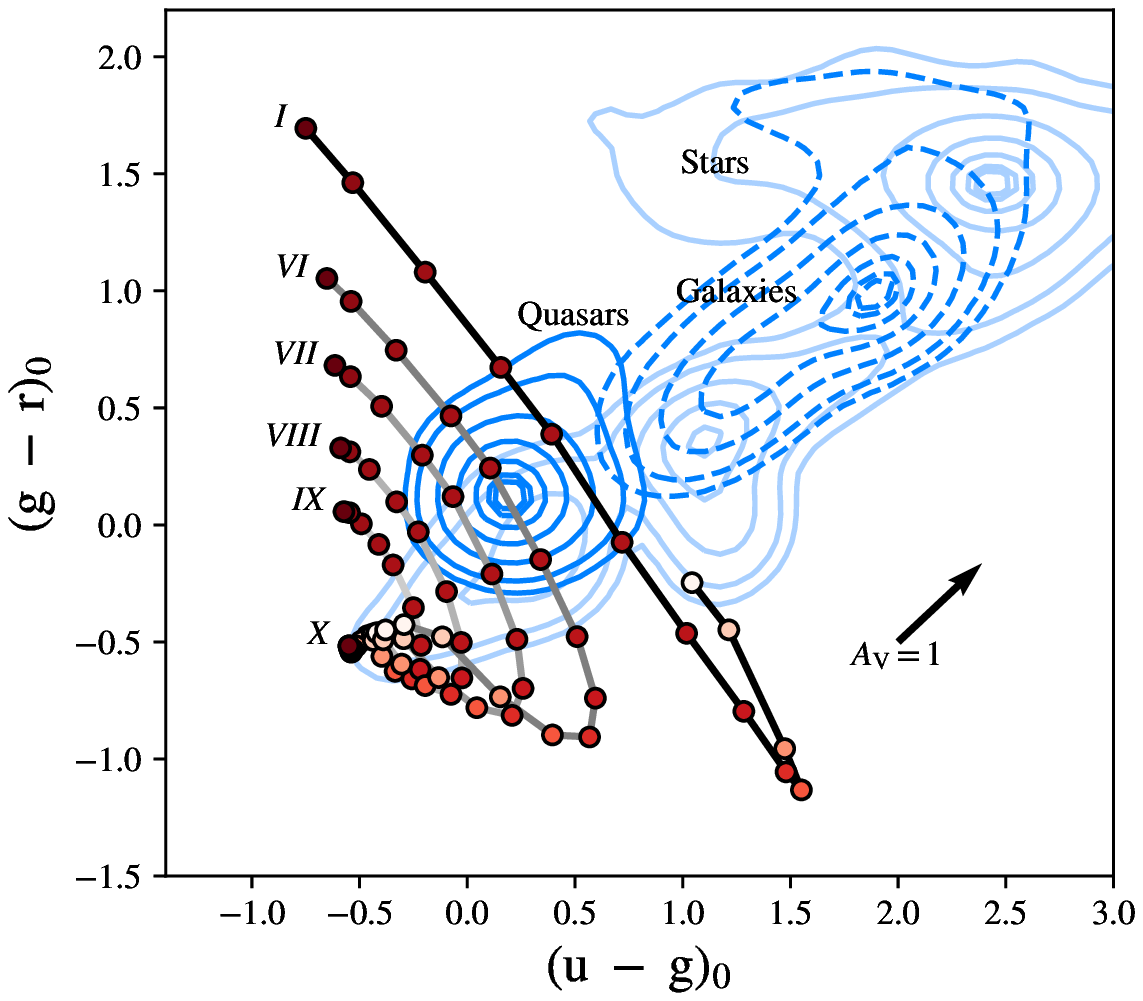} 
\includegraphics[width=3in]{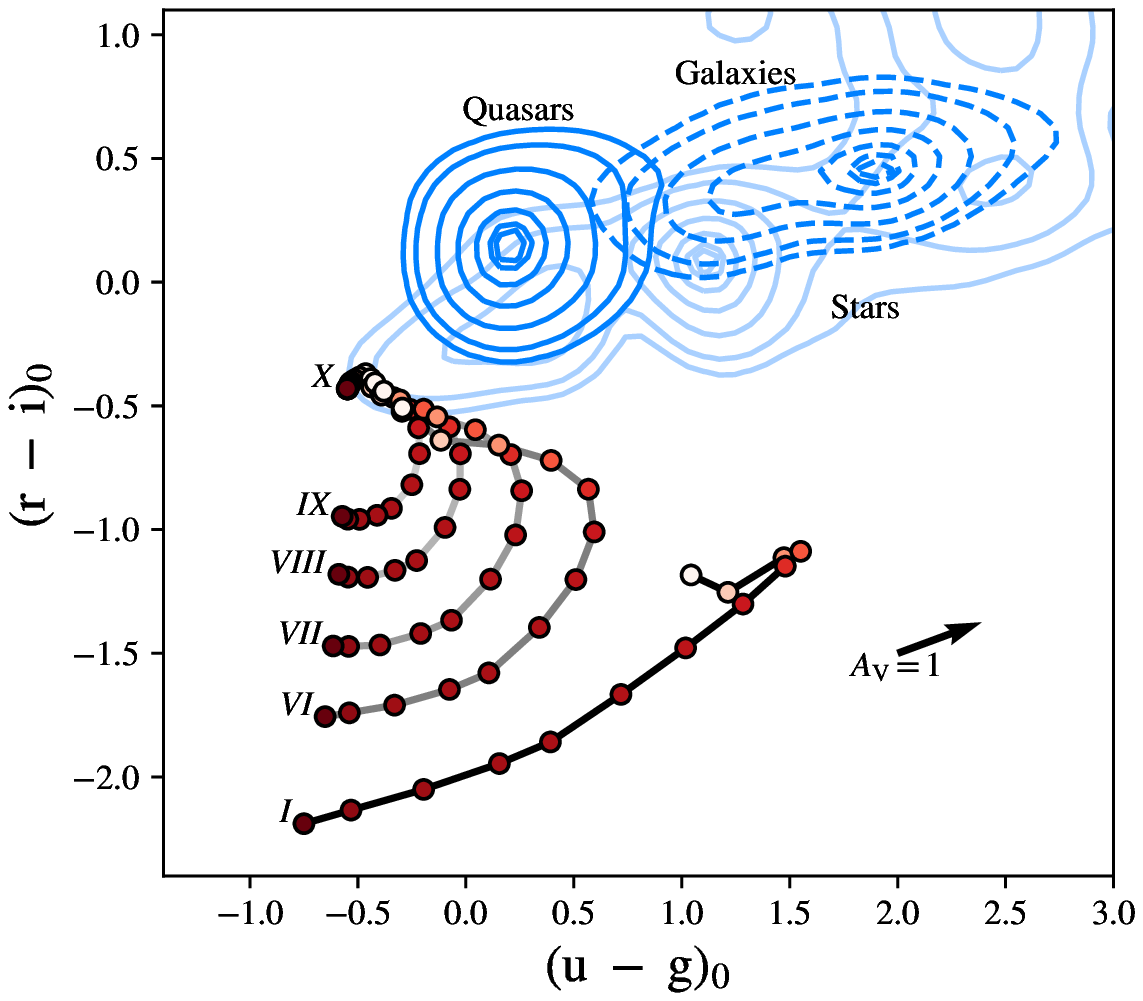}
\caption[Color-Color Diagrams]{Selected color-color diagrams for PN models along with distributions of point sources from SDSS. 
Points and grey-scaled lines are described in Figure \ref{fig:cmd}.
Distributions of SDSS objects have been smoothed with a Gaussian filter with a $\sigma=0.13$~mag.
The contour levels indicate 5, 10, 25, 50, 75, 90, and 95\% of the maximum value of the smoothed distributions.
}
\label{fig:colorcolor}
\end{figure}

We compare our PN models and variants to SDSS objects in Figure~\ref{fig:colorcolor}.
Contours of stars, galaxies, and quasars are used to illustrate their locations in the various diagrams. 
Data for these object types come from SDSS DR7 \citep{2009ApJS..182..543A} and were queried using the SDSS SkyServer. 
We only included objects with SDSS spectroscopic observations to allow for further filtering on $z$ for the star, galaxy, and quasar object types.

The top panel of Figure~\ref{fig:colorcolor} shows the $r-i$ vs $g-r$ color-color diagram.
Here, models occupy a separated space from the SDSS data with $r-i$ between 0 and $-2.5$ and $g-r$ between $\sim$1.75 and $-1.5$.
The unresolved synthetic nebular models (variant I), as well as the oldest resolved models (variants VI through IX) remain separated from stars, quasars, and galaxies through out all or most stages of evolution in this diagram for $r-i < -0.75$. 
(We do not show variants II--V as these are effectively equivalent to variant I.)
As the central star and nebula are resolved, the colors converge towards the locus of blue evolved stars, as expected.
Similarly, the bottom panel of Figure~\ref{fig:colorcolor} shows that our models are also separated from most of the SDSS data in the $r-i$ vs $u-g$ color-color diagram for $r-i < -0.75$.
These two diagrams will be essential in identifying new PN candidates of all ages.

In the middle panel of Figure \ref{fig:colorcolor}, the PN models occupy the left region of the diagram.
They fall between values of $\sim$1.75 to $-1.5$ in $g-r$ and $\sim$1.6 to $-1.0$ in $u-g$.
Our synthetic nebulae cross through the loci of early-type MS stars, quasars, and blue evolved stars in this diagram.
The youngest ($t_{\rm age} < 7000$ yrs) and oldest ($t_{\rm age} > 7500$ yrs) stages in our fully unresolved models (variant I) are somewhat separated from these loci making these areas good places to look for young and old unresolved PN candidates.
The change in color values for all CCDs is a result of the nebular emission as seen in Figure~\ref{fig:cmd} and is described in the previous section.

\section{Discussion}\label{sec:discussion}

\subsection{Comparison to Prior Work}

A number of studies have shown that emission line objects like PNe can be identified or characterized by broadband filters.
In the UV, optical, and near-infrared (NIR) \citet{2014ApJ...792..121V} used Hubble Space Telescope (HST) Advanced Camera for Surveys and the Panchromatic Hubble Andromeda Treasury (PHAT) survey data to identify broadband detections of known PNe and were able to roughly estimate the excitation classifications of groups of PNe with the addition of archival $\textit{m}$5007 narrow-band magnitudes.
In the optical, \citet{2014AJ....147...16K} used the Sloan Digital Sky Survey (SDSS) colors to separate PNe from other point sources in the outskirts of M31.
\citet{2016JPhCS.728c2008P} developed a method using mid-infrared colors to identify galactic PN candidates using available data in the Galactic Legacy Infrared Mid-Plane Survey Extraordinaire I (GLIMPSE I) point source archive.
Data from the Infrared Astronomical Satellite (IRAS) was used to inform optical spectroscopic follow-up by \citet{2006AA...458..173S} to confirm PNe candidates along with post-AGB stars and sources transitioning from AGB to PN stages of evolution.
\citet{2008AA...480..409C} and following papers in the series \citep{2010AA...509A..41C,2014AA...567A..49R} describe a similar method for selecting symbiotic star candidates using the INT Photometric $\rm H\alpha$ Survey (IPHAS) and the Two Micron All Sky Survey (2MASS) colors showing that symbiotic stars can be distinguished from PNe through colors but ultimately spectroscopic follow-up is necessary for verification.

\begin{figure*}[!ht]
\centering
\includegraphics[width=3.5in]{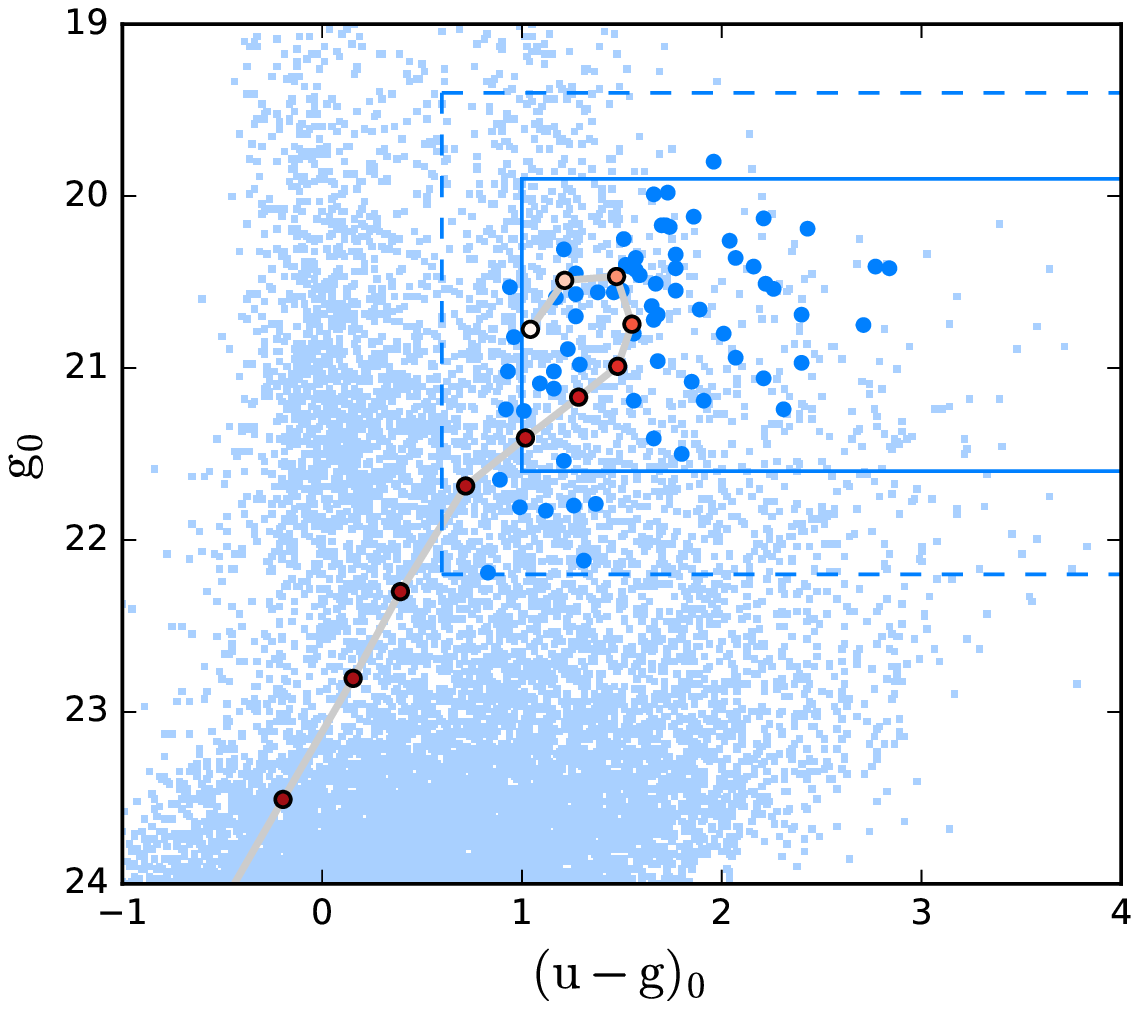}
\includegraphics[width=3.5in]{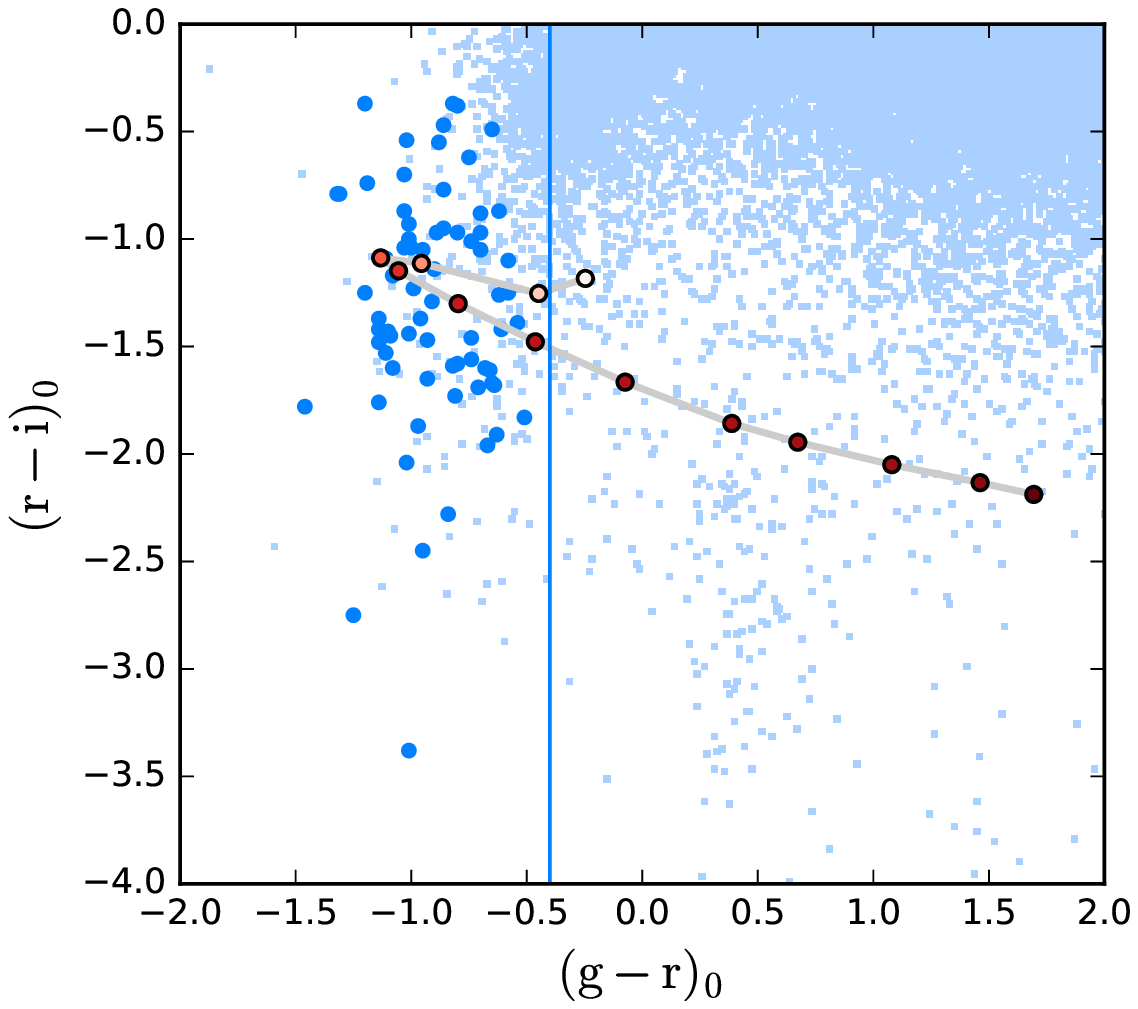}
\caption[M31 PNe]{70 confirmed PNe, in dark blue, detected at the outskirts of the Andromeda galaxy (M31) as point sources by SDSS \citep{2014AJ....147...16K}. PNe models are colored similarly to Figure~\ref{fig:HR}. Left panel shows some of the color cuts used by \citet{2014AJ....147...16K} as solid and dashed lines at $\rm(u - g)_{0} \geq 1.0~mags$ and 0.6 mags, $\rm 19.9 \geq g_{0} \leq 21.6~mags$ and $\rm 19.4 \geq g_{0} \leq 22.4~mags$. Right panel shows a color-magnitude diagram of the same objects with solid line showing the $\rm (g - r \geq -0.4)$ color cut. SDSS stars near M31 are shown in light blue.}
\label{fig:Kniazev pne}
\end{figure*}

We compare our predicted PNe colors and magnitudes (for variant I) to confirmed PNe in M31 \citep{2014AJ....147...16K}. 
\citet{2014AJ....147...16K} used 29 known PNe in M31 \citep{1986ApJ...304..490J,1987ApJ...317...62N} to construct magnitude and color criteria to identify additional PNe with SDSS photometry. 
70 were confirmed to be PNe through follow-up spectroscopy \citep{2014AJ....147...16K}.
In Figure~\ref{fig:Kniazev pne} the color-magnitude and color-color diagrams presented in \citet{2014AJ....147...16K} are reproduced along with some of their selection criteria. 
We queried SDSS \citep[DR14;][]{2018ApJS..235...42A} for clean sources within 1 degree of M31 and with point source criteria outlined in \citet{2014AJ....147...16K}.
The spectroscopically verified PNe from \citet{1987ApJ...317...62N} along with our 3 $M_{\odot}$ PN evolutionary models scaled to the distance of M31 \citep[760 kpc;][]{1999AARv...9..273V} are overlaid onto the CMD and CCD.
At such a distance, all of our nebular models are unresolved. 
In the CMD, the sample identified by \citet{2014AJ....147...16K} is consistent with the youngest stages of our PN models ($t_{\rm age}<7300$~years). 
We find similar consistency in the CCD, except for youngest stage ($t_{\rm age}\sim4200$~years), which is outside of the $\rm (g - r \geq -0.4)$ color cut used by \citet{2014AJ....147...16K} to reduce contamination from high redshift objects. 
As suggested by our PN models, this color cut will limit the identification of the youngest and oldest PNe in M31, however, these PNe might be present and identifiable using the CCD. 

\begin{figure*}[!ht]
\centering
\includegraphics[width=3.5 in]{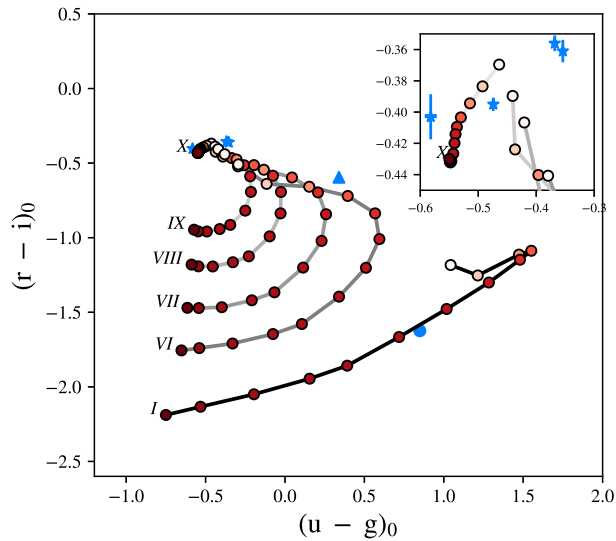}
\caption[yuan pne]{Color-color diagram showing PNe models and relevant variants. 
Red points connected with grey-scaled lines are described in Figure \ref{fig:cmd}.
Blue points of different marker types are faint PNe from  \citet{2013MNRAS.436..718Y}, color values are from SDSS and include error bars.
Blue marker's correspond to different PNe sizes; circle $\leq$ $3\farcs0$, $3\farcs0$ $<$ triangle $<$ $20\farcs0$, stars $\geq$ $20\farcs0$.
Variants labeled with higher numerals contain less nebular emission, representing resolved PNe. Inset plot shows region near variant X for clarity.}
\label{fig:Yuan pne}
\end{figure*}

We also compared our results for resolved and unresolved colors and magnitudes to six faint PNe (PN G094.0+27.4, PN G211.4+18.4, PN G049.3+88.1, PN G025.3+40.8, PN G047.0+42.4, and PN G158.8+37.1) recovered by \citet{2013MNRAS.436..718Y} through identification of excess $\rm[OIII]~\lambda5007$ or $\rm[OIII]~\lambda4959$ emission in the SDSS spectra of objects who's photometry was affected by the nebula. 
For each PN, we used the object's coordinates to obtain an SDSS g band image of the region \citep[DR14;][]{2018ApJS..235...42A} and visually identified the PN in the image (using ds9).
We then visually located the closest object to the CS position of the PN and obtained the SDSS ugriz magnitudes of that source.
In Figure~\ref{fig:Yuan pne} we compare the colors of these six sources in relation to our models. 
The colors for PN G094.0+27.4, PN G211.4+18.4, PN G047.0+42.4, and PN G158.8+37.1 are consistent with that expected from a central star with very little nebular emission as they appear close to our model variants VIII-X (marked as stars in Figure~\ref{fig:Yuan pne}).
The nebular radii of these three objects are \textgreater\ $90^{\prime\prime}$, which is much larger than the typical $\sim1^{\prime\prime}$ seeing resolution limit of SDSS, hence only a small fraction of the nebular emission is detected in these objects. 
PN G025.3+40.8 (marked as a triangle in Figure~\ref{fig:Yuan pne}), with a nebular radius of $13^{\prime\prime}$, is consistent with a higher fraction of nebular flux detected as it appears near our variant VI models indicating a fractional nebular flux of $\sim9\times10^{-3}$ was observed with the CS.
PN G049.3+88.1 (marked as a circle in Figure~\ref{fig:Yuan pne}), with a nebular radius of $2\farcs7$, is consistent with nearly all of the nebular flux detected as it appears near our variant I models.
While this is not a rigorous comparison of these sources to our models, generally, we see that CSPNe with a larger sized nebula contain less nebular flux than CSPNe with a smaller sized nebula as predicted by our models.

\subsection{Applications to Upcoming Surveys}
\hspace*{3mm}
Upcoming surveys such as LSST can potentially greatly expand the number of known PNe at different stages of evolution, both within the Milky Way and in other nearby galaxies. As an example, other authors have estimated LSST's distance limit for RR Lyrae variables to be 400~kpc  \citep{2009arXiv0912.0201L}; LSST's potential reach for PNe may also be similarly impressive.
LSST will commence operations in 2023 and is scheduled to be a 10-year long survey of the southern sky aiming at addressing questions from scales of the solar system to dark energy.
The telescope harbors an 8.4 meter mirror and a 3,200 megapixel camera that will image 37 billion stars and galaxies.
The camera is equipped with the $ugrizy$ filter set with sensitivity information detailed in Table~\ref{table: Filter Properties}.

\begin{figure*}[!ht]
\centering
\includegraphics[width = 5.5 in]{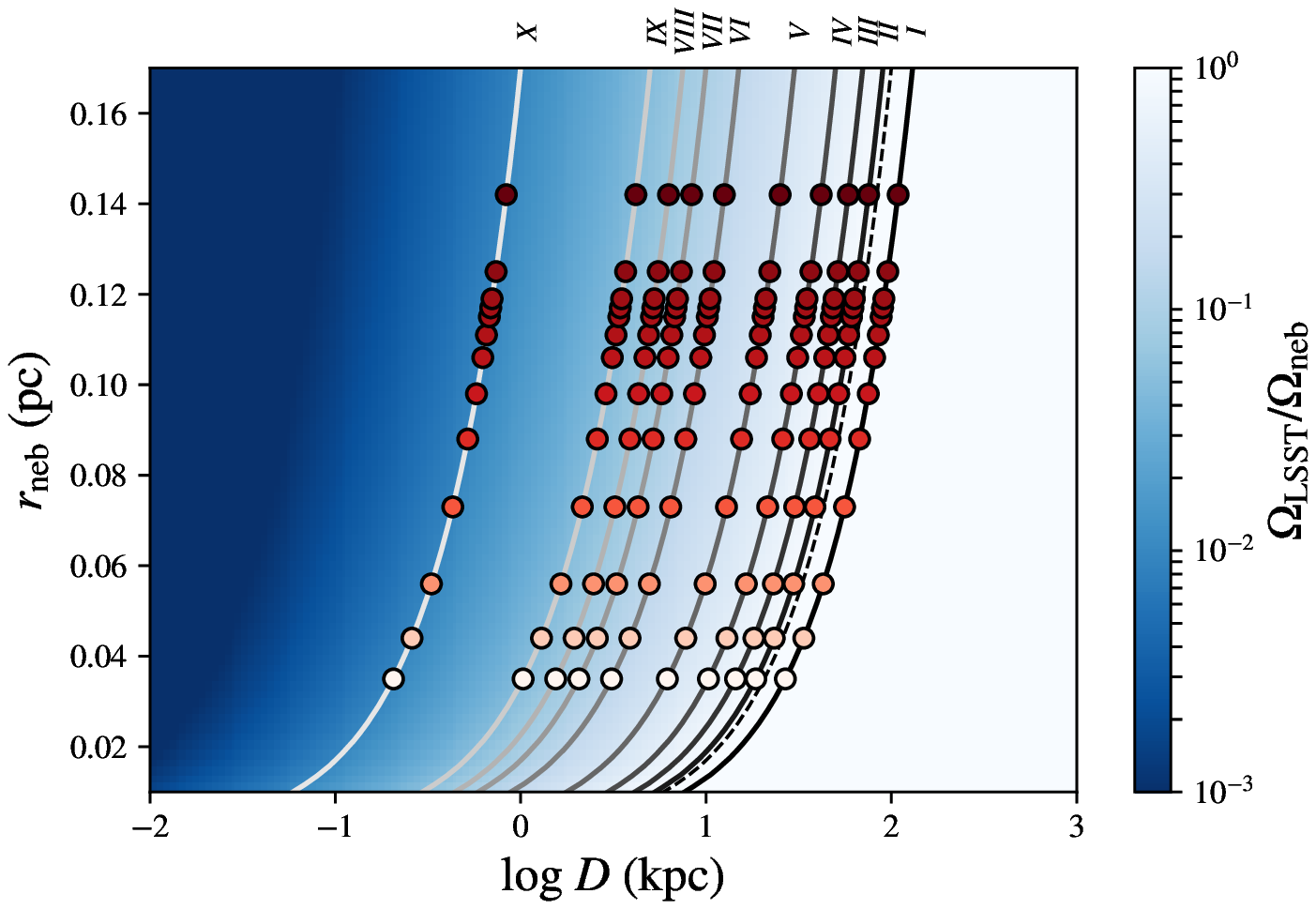}
\caption[Resolution Diagram]{PNe model radius (pc) versus PNe distance is shown, along with the degree to which the PN is resolved by the nominal LSST seeing ($0\farcs7$) relative to the PN radius ($\Omega_{\rm LSST}/\Omega_{\rm neb}$) in the blue shading. 
Points and grey-scaled lines are described in Figure \ref{fig:cmd}, Roman numerals label fractional variants.
The point where LSST begins to resolve the PNe, $\Omega_{\rm LSST}/\Omega_{\rm neb} = 1$, falls between variants I and II (dashed line). PNe at larger distances or with smaller radii will be unresolved to LSST;
PNe at smaller distances or larger radii will be resolved to a degree depending on these factors.
}
\label{fig:distancelimit}
\end{figure*}

The first consideration for LSST's reach of PNe is spatial resolution of PNe flux. 
The optimal seeing limit for LSST is projected to be $\sim0\farcs7$ \citep{2009arXiv0912.0201L}, if we take this as an estimate of the size of aperture ($\Omega_{\rm LSST}$), then for any $\Omega_{\rm neb}>\Omega_{\rm LSST}$ the nebula will potentially be resolved. 
In Figure~\ref{fig:distancelimit}, we calculate the ratio $\Omega_{\rm LSST}/\Omega_{\rm neb}$ for a range of distances and include the locations of our synthetic nebular models shown in Figures~\ref{fig:cmd}~and~\ref{fig:colorcolor}. 
At a distance to the Magellanic Clouds \citep[49.97 and 62.1 kpc;][]{2013Natur.495...76P,2014ApJ...780...59G} young models ($\rm t_{\rm age} < 6300~yrs$) are unresolved while older models are marginally resolved. 
For our oldest models ($\rm t_{\rm age} > 6300~yrs$) the nebular diameters are $\sim1\farcs0$. 
With worse seeing or by utilizing a slightly adjustable aperture in the imaging analysis, it is possible to observe all the nebular flux for marginally-resolved cases. 

\begin{figure}[!ht]
\centering
\includegraphics[width=3in]{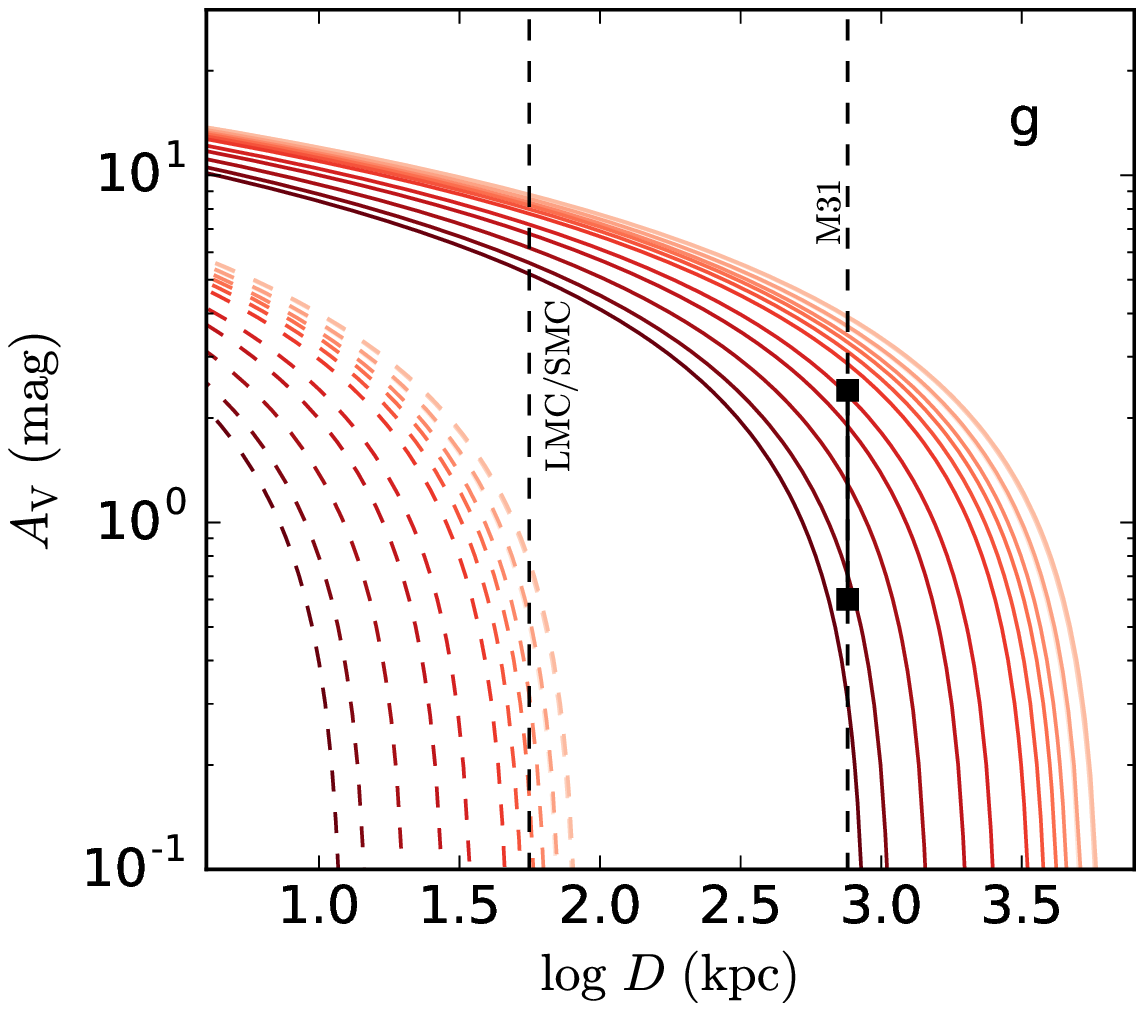}
\includegraphics[width=3in]{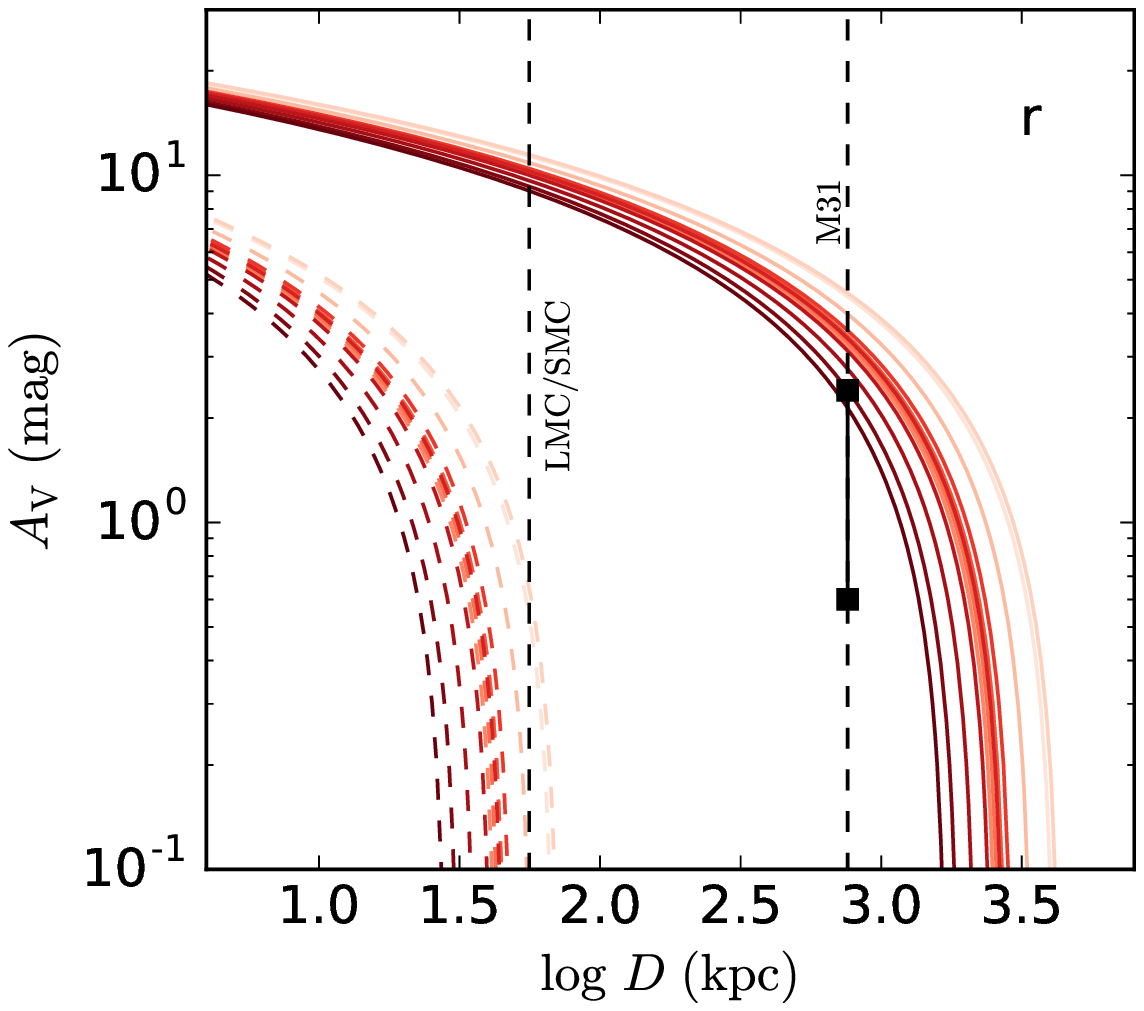}
\includegraphics[width=3in]{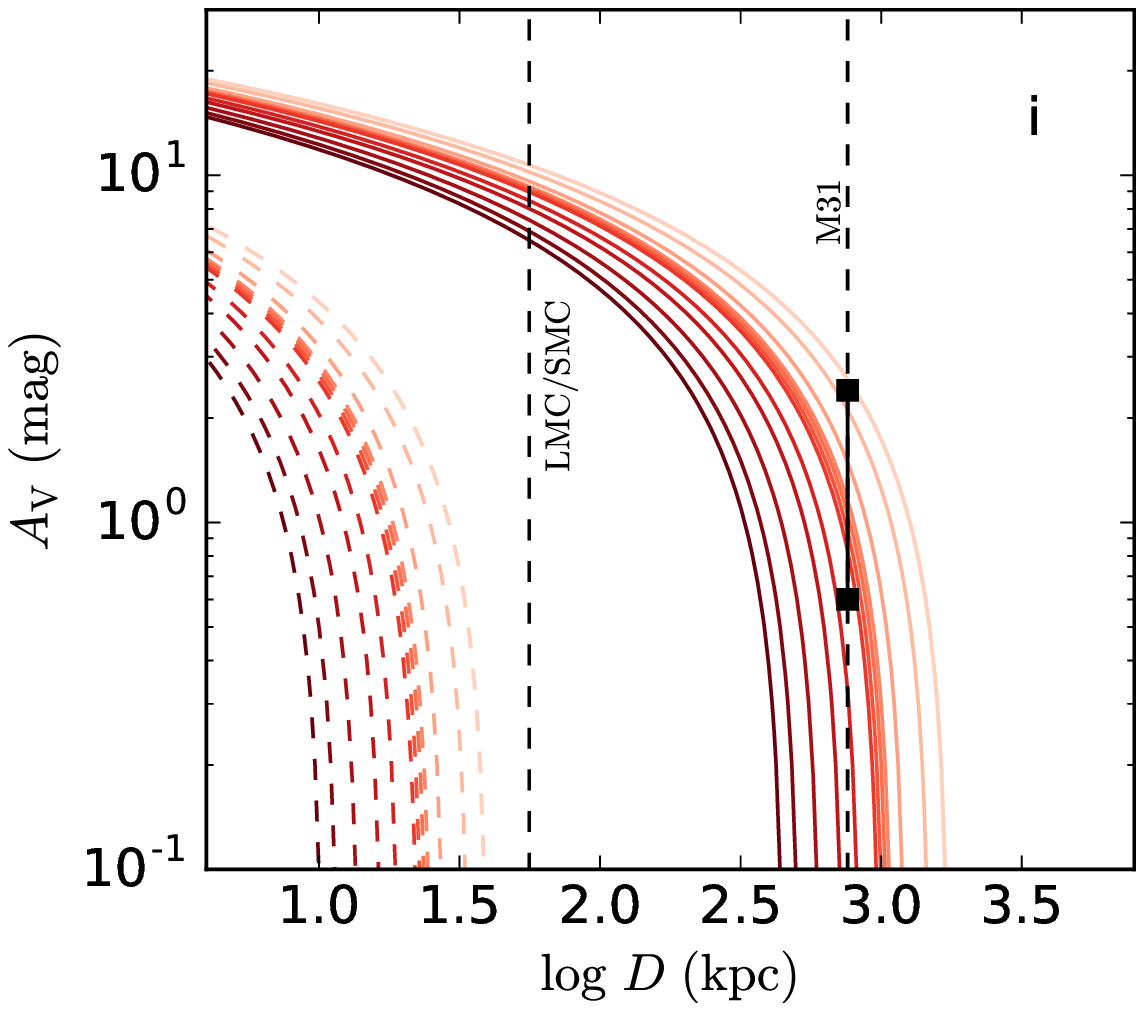}
\caption[Magnitude Limits]{Magnitude limits based on the LSST saturation and 5$\sigma$ detection limits for three filters as a function of distance and extinction \citep{2009arXiv0912.0201L}.
Models shown represent fully unresolved PNe.
Lines shaded for age, lighter correspond to younger models.
Dashed red lines show distance at which PNe will saturate for given amount of extinction.
Solid lines show distance at which PNe will reach the 5$\sigma$ detection limit.
Black squares show range of extinction values for M31 ($0.6 \geq A_{\rm V} \leq 2.4$ \citep{2015ApJ...814....3D})}
\label{fig:maglimit}
\end{figure}

The reach for a given survey is also dependent on the survey saturation and limiting magnitudes \citep[see Table \ref{table: Filter Properties};][]{2009arXiv0912.0201L}, as well as extinction to the source. 
Using Table~\ref{table: PNe Magnitudes}, it is possible to study unresolved nebular fluxes for specific combinations of distance and extinction to determine the reach for LSST. 
Because the identification of potential PNe requires detection in at least three filters ($g$, $r$, and $i$) we considered the saturation and 5$\sigma$ limiting magnitudes, $m_{\rm sat}$ and $m_{5\sigma}$, respectively, for these three filters.
In Figure~\ref{fig:maglimit}, we determined the LSST detection limit for our unresolved nebular models as a function of distance and extinction. 
For modest extinction values ($A_{\rm V} < \rm 5~mag$) at the LMC/SMC, all of our nebular models predict detectable emission in $g$, $r$, and $i$, however, we potentially reach the saturation limit in $g$ and $r$, for our youngest models when extinction drops below 1 mag (see Figure~\ref{fig:maglimit}). 
For another example, if LSST observed a system like Andromeda (M31) at a distance of 760 kpc \citep{1999AARv...9..273V}, in the areas of low extinction ($A_{\rm V}\lesssim 2{\rm ~mag}$), similar to those found in the outskirts of that galaxy \citep{2014AJ....147...16K, 2015ApJ...814....3D}, the youngest nebular models can be detected in these three bands, but fewer of the older nebular models would be detected in $i$ ($\rm t_{\rm age} > 7300~yrs$). 
Hence, surveys like LSST can be used to discover many new potential PNe beyond our galaxy. 

We can also consider LSST's reach for PNe specifically within the Milky Way.
Here, estimating the reach is further complicated by the numerous combinations of nebular age/size, distance, and extinction.
The contribution of the nebular flux to the aperture is a strong function of $\Omega_{\rm LSST}/\Omega_{\rm neb}$ (see Figure~\ref{fig:cmd}). 
To consider the reach within the Milky Way, we provide variants of the absolute magnitudes in Table~\ref{table: neb_fraction} for a range of $\Omega_{\rm LSST}/\Omega_{\rm neb}$.
For example, we considered the distance-$A_{\rm V}$ based reach for LSST at a distance consistent with the galactic center \citep[8.3 kpc;][]{2018ApSS.363..127M}.  
Young PNe ($t_{\rm age} \approx 3500{\rm~yr}$, $r_{\rm neb}\sim0.035{\rm ~pc}$) will be resolved at the galactic center with only $\sim$1.6\% of the nebular emission measured by a $0\farcs7$ aperture ($\Omega_{\rm LSST}/\Omega_{\rm neb}=0.075$). 
As a result, such a PN would be detectable in $g$, $r$, and $i$ for lines of sight where $A_{\rm V}<10{\rm ~mags}$. 
However, for lines of sight where $A_{\rm V}$ drops below $2{\rm ~mags}$, the model-predicted flux would reach the saturation limits. 
For older models ($\rm t_{\rm age} > 6300~yrs$) at the distance of the galactic center, since the nebula grows in size, the nebular flux contribution reduces further.
For these older models, saturation is unlikely with $A_V$ serving as the only limit on detection with maximum line of sight $A_V$ values in the range of $5-10{\rm ~mags}$. 

\subsection{Limitations of the Current Work}
\hspace*{3mm}
The methodology and results presented in this work demonstrate the potential of using synthetic optical PNe spectra and a set of filters to produce accurate magnitudes and colors. 
However, we have only considered the evolutionary track for a $\rm 3~M_{\odot}$ progenitor with chemical abundances similar to those of galactic PNe.
Constructing these models for a range of progenitor masses will give a better understanding of the distribution of PNe in color-color space and their identification in future surveys.

In developing our methodology, we discovered that a crucial ingredient to construct accurate models is the hydrodynamical expansion over time of the nebular radius as the central star evolves. 
In particular, self-consistent radiation-hydrodynamic models of a PN \citep[e.g.,][]{2013AA...558A..78J,2005AA...441..573S,2014MNRAS.443.3486T} are important ingredients necessary to track to evolutionary behavior of the nebula and its radiation. 

As mentioned in section 2.4, this method can only identify PNe candidates. 
All potential PNe candidates will require supporting spectroscopic or narrowband observations to verify their nature. 
This is also necessary to distinguish true PNe from other emission line objects such as Wolf-Rayet (WR) stars, HII regions, symbiotic stars, and CVs which may contaminate the same color-color space.

Finally, our study highlights an immediate outcome and discovery space made possible by a survey like LSST. 
We have not carefully considered the impact of repeated observations, which can be used to search for variability of the central star and/or potential companions and push the 5-$\sigma$ limiting magnitude fainter when observations are combined. 

\section{Conclusions}\label{sec:summary}
\hspace*{3mm}

In this work we presented synthetic absolute magnitudes in the $ugrizy$ filters of 13 PN models representing the evolution of a PN for a $\rm3~M_{\odot}$ progenitor star over $\sim$5,000 years (Table \ref{table: PNe Magnitudes}).
We have calculated the colors for various photometric aperture sizes to explore spatially resolved and unresolved cases.
We showed that our model magnitudes and colors are consistent with real observations for both resolved and unresolved cases.
We also showed that LSST will allow for the identification of many PNe in the Milky Way and neighboring galaxies.

Color-color diagrams will be useful in differentiating PNe from other point sources in the upcoming LSST era.
We showed that the PNe model colors are entirely separate from most other SDSS objects cataloged in SDSS;
indeed, we can simply provide a cutoff above $r - i = -0.75$ to find viable PNe candidates.
Moreover, we showed that colors can also differentiate younger PNe ($\rm t_{\rm age} < 7000~yrs$) from older PNe ($\rm t_{\rm age} > 7500~yrs$).
This will likely still be possible even in the presence of large amounts of extinction.
There is still work to be done on how colors of PNe with various progenitor masses will change.

Given the LSST seeing limited resolution and magnitude limits, PNe detected at various distances will be spatially resolved to varying degrees.
For distances to the galactic center, the youngest PNe ($\rm t_{\rm age} \approx 3500~yrs$) will saturate with $\rm A_{\rm V} \leq 2~mags$, be visible up to $\rm A_{\rm V} \approx 10~mags$, and be partially resolved.
The oldest PNe ($\rm t_{\rm age} \approx 8300~yrs$) will be visible up to $\rm A_{\rm V} \approx 5~mags$ and be mostly resolved.
Young PNe ($\rm t_{\rm age} < 6300~yrs$) at the LMC/SMC will be fully unresolved and visible for moderate $\rm A_{\rm V}$ between $\rm \sim1~and~5~mags$ while older models will be resolved containing at least 20\% of the nebular emission.
For distances similar to M31, all PNe will be unresolved but only younger PNe ($\rm t_{\rm age} < 7300~yrs$) will be visible with $\rm A_{\rm V}$ between $\rm \sim0.6~and~2.4~mags$.
This and future works will help prepare the astronomical community for the massive amounts of data that will be provided by LSST and enable the discovery of unprecedented numbers of new PNe, enabling detailed tests of our current theories of this phase of stellar evolution.

\section*{Acknowledgements}
\hspace*{3mm}
The specific use of Cloudy in this work is through the Python wrapper pyCloudy which allows the user to interact with the Cloudy input and output files through Python \citep{2013ascl.soft04020M}.
This research has made use of the Spanish Virtual Observatory (http://svo.cab.inta-csic.es) supported from the Spanish MINECO/FEDER through grant AyA2014-55216.
This research has made use of the VizieR catalogue access tool, CDS, Strasbourg, France.
This research has made use of the SIMBAD database, operated at CDS, Strasbourg, France.
This work made use of the IPython package \citep{2007CSE.....9c..21P}.
This research made use of SciPy \citep{jones_scipy_2001}.
This research made use of NumPy \citep{van2011numpy} This research made use of matplotlib, a Python library for publication quality graphics \citep{Hunter:2007}.
This research made use of ds9, a tool for data visualization supported by the Chandra X-ray Science Center (CXC) and the High Energy Astrophysics Science Archive Center (HEASARC) with support from the JWST Mission office at the Space Telescope Science Institute for 3D visualization. 
This research made use of Astropy, a community-developed core Python package for Astronomy \citep{2013AA...558A..33A}.
Funding for the Sloan Digital Sky Survey IV has been provided by the Alfred P. Sloan Foundation, the U.S. Department of Energy Office of Science, and the Participating Institutions.
SDSS-IV acknowledges support and resources from the Center for High-Performance Computing at the University of Utah.
The SDSS web site is www.sdss.org.
SDSS-IV is managed by the Astrophysical Research Consortium for the Participating Institutions of the SDSS Collaboration including the Brazilian Participation Group, the Carnegie Institution for Science, Carnegie Mellon University, the Chilean Participation Group, the French Participation Group, Harvard-Smithsonian Center for Astrophysics, Instituto de Astrof\'isica de Canarias, The Johns Hopkins University, Kavli Institute for the Physics and Mathematics of the Universe (IPMU) / University of Tokyo, Lawrence Berkeley National Laboratory, Leibniz Institut f\"ur Astrophysik Potsdam (AIP), Max-Planck-Institut f\"ur Astronomie (MPIA Heidelberg), Max-Planck-Institut f\"ur Astrophysik (MPA Garching), Max-Planck-Institut f\"ur Extraterrestrische Physik (MPE), National Astronomical Observatories of China, New Mexico State University, New York University, University of Notre Dame, Observat\'ario Nacional / MCTI, The Ohio State University, Pennsylvania State University, Shanghai Astronomical Observatory, United Kingdom Participation Group, Universidad Nacional Aut\'onoma de M\'exico, University of Arizona, University of Colorado Boulder, University of Oxford, University of Portsmouth, University of Utah, University of Virginia, University of Washington, University of Wisconsin, Vanderbilt University, and Yale University.
Funding for the SDSS and SDSS-II has been provided by the Alfred P. Sloan Foundation, the Participating Institutions, the National Science Foundation, the U.S. Department of Energy, the National Aeronautics and Space Administration, the Japanese Monbukagakusho, the Max Planck Society, and the Higher Education Funding Council for England.
The SDSS is managed by the Astrophysical Research Consortium for the Participating Institutions. The Participating Institutions are the American Museum of Natural History, Astrophysical Institute Potsdam, University of Basel, University of Cambridge, Case Western Reserve University, University of Chicago, Drexel University, Fermilab, the Institute for Advanced Study, the Japan Participation Group, Johns Hopkins University, the Joint Institute for Nuclear Astrophysics, the Kavli Institute for Particle Astrophysics and Cosmology, the Korean Scientist Group, the Chinese Academy of Sciences (LAMOST), Los Alamos National Laboratory, the Max-Planck-Institute for Astronomy (MPIA), the Max-Planck-Institute for Astrophysics (MPA), New Mexico State University, Ohio State University, University of Pittsburgh, University of Portsmouth, Princeton University, the United States Naval Observatory, and the University of Washington. 

George Vejar thanks the LSSTC Data Science Fellowship Program, his time as a Fellow has benefited this work.

\bibliography{Mendeley.bib}

\begin{thebibliography}{}
\expandafter\ifx\csname natexlab\endcsname\relax\def\natexlab#1{#1}\fi

\bibitem[{Abazajian {et~al.}(2009)Abazajian, Adelman-McCarthy, Ag{\"{u}}eros,
  Allam, Prieto, An, Anderson, Anderson, Annis, Bahcall, Bailer-Jones,
  Barentine, Bassett, Becker, Beers, Bell, Belokurov, Berlind, Berman,
  Bernardi, Bickerton, Bizyaev, Blakeslee, Blanton, Bochanski, Boroski,
  Brewington, Brinchmann, Brinkmann, Brunner, Budav{\'{a}}ri, Carey, Carliles,
  Carr, Castander, Cinabro, Connolly, Csabai, Cunha, Czarapata, Davenport,
  de~Haas, Dilday, Doi, Eisenstein, Evans, Evans, Fan, Friedman, Frieman,
  Fukugita, G{\"{a}}nsicke, Gates, Gillespie, Gilmore, Gonzalez, Gonzalez,
  Grebel, Gunn, Gy{\"{o}}ry, Hall, Harding, Harris, Harvanek, Hawley, Hayes,
  Heckman, Hendry, Hennessy, Hindsley, Hoblitt, Hogan, Hogg, Holtzman, Hyde,
  Ichikawa, Ichikawa, Im, Ivezi{\'{c}}, Jester, Jiang, Johnson, Jorgensen,
  Juri{\'{c}}, Kent, Kessler, Kleinman, Knapp, Konishi, Kron, Krzesinski,
  Kuropatkin, Lampeitl, Lebedeva, Lee, Lee, Leger, L{\'{e}}pine, Li, Lima, Lin,
  Long, Loomis, Loveday, Lupton, Magnier, Malanushenko, Malanushenko,
  Mandelbaum, Margon, Marriner, Mart{\'{i}}nez-Delgado, Matsubara, McGehee,
  McKay, Meiksin, Morrison, Mullally, Munn, Murphy, Nash, Nebot, Neilsen,
  Newberg, Newman, Nichol, Nicinski, Nieto-Santisteban, Nitta, Okamura,
  Oravetz, Ostriker, Owen, Padmanabhan, Pan, Park, Pauls, Peoples, Percival,
  Pier, Pope, Pourbaix, Price, Purger, Quinn, Raddick, Fiorentin, Richards,
  Richmond, Riess, Rix, Rockosi, Sako, Schlegel, Schneider, Scholz, Schreiber,
  Schwope, Seljak, Sesar, Sheldon, Shimasaku, Sibley, Simmons, Sivarani, Smith,
  Smith, Smol{\v{c}}i{\'{c}}, Snedden, Stebbins, Steinmetz, Stoughton, Strauss,
  SubbaRao, Suto, Szalay, Szapudi, Szkody, Tanaka, Tegmark, Teodoro, Thakar,
  Tremonti, Tucker, Uomoto, Vanden~Berk, Vandenberg, Vidrih, Vogeley, Voges,
  Vogt, Wadadekar, Watters, Weinberg, West, White, Wilhite, Wonders, Yanny,
  Yocum, York, Zehavi, Zibetti, \& Zucker}]{2009ApJS..182..543A}
Abazajian, K.~N., Adelman-McCarthy, J.~K., Ag{\"{u}}eros, M.~A., {et~al.} 2009,
  The Astrophysical Journal Supplement Series, 182, 543

\bibitem[{Abolfathi {et~al.}(2018)Abolfathi, Aguado, Aguilar, Prieto, Almeida,
  Ananna, Anders, Anderson, Andrews, Anguiano, Arag{\'{o}}n-Salamanca,
  Argudo-Fern{\'{a}}ndez, Armengaud, Ata, Aubourg, Avila-Reese, Badenes,
  Bailey, Balland, Barger, Barrera-Ballesteros, Bartosz, Bastien, Bates,
  Baumgarten, Bautista, Beaton, Beers, Belfiore, Bender, Bernardi, Bershady,
  Beutler, Bird, Bizyaev, Blanc, Blanton, Blomqvist, Bolton, Boquien,
  Borissova, Bovy, Bradna~Diaz, Nielsen~Brandt, Brinkmann, Brownstein, Bundy,
  Burgasser, Burtin, Busca, Ca{\~{n}}as, Cano-D{\'{i}}az, Cappellari, Carrera,
  Casey, Sodi, Chen, Cherinka, Chiappini, Choi, Chojnowski, Chuang, Chung,
  Clerc, Cohen, Comerford, Comparat, do~Nascimento, da~Costa, Cousinou, Covey,
  Crane, Cruz-Gonzalez, Cunha, Ilha, Damke, Darling, Davidson, Dawson,
  de~Icaza~Lizaola, Macorra, de~la Torre, De~Lee, Sainte~Agathe,
  Deconto~Machado, Dell’Agli, Delubac, Diamond-Stanic, Donor, Downes, Drory,
  Mas~des Bourboux, Duckworth, Dwelly, Dyer, Ebelke, Eigenbrot, Eisenstein,
  Elsworth, Emsellem, Eracleous, Erfanianfar, Escoffier, Fan, Alvar,
  Fernandez-Trincado, Cirolini, Feuillet, Finoguenov, Fleming, Font-Ribera,
  Freischlad, Frinchaboy, Fu, Chew, Galbany, Garc{\'{i}}a~P{\'{e}}rez,
  Garcia-Dias, Garc{\'{i}}a-Hern{\'{a}}ndez, Garma~Oehmichen, Gaulme, Gelfand,
  Gil-Mar{\'{i}}n, Gillespie, Goddard, Gonz{\'{a}}lez~Hern{\'{a}}ndez,
  Gonzalez-Perez, Grabowski, Green, Grier, Gueguen, Guo, Guy, Hagen, Hall,
  Harding, Hasselquist, Hawley, Hayes, Hearty, Hekker, Hernandez,
  Hernandez~Toledo, Hogg, Holley-Bockelmann, Holtzman, Hou, Hsieh, Hunt,
  Hutchinson, Hwang, Jimenez~Angel, Johnson, Jones, J{\"{o}}nsson, Jullo,
  Sakil~Khan, Kinemuchi, Kirkby, Kirkpatrick~IV, Kitaura, Knapp, Kneib,
  Kollmeier, Lacerna, Lane, Lang, Law, Le~Goff, Lee, Li, Li, Lian, Liang, Lima,
  Lin, Long, Lucatello, Lundgren, Mackereth, MacLeod, Mahadevan, Geimba~Maia,
  Majewski, Manchado, Maraston, Mariappan, Marques-Chaves, Masseron, Masters,
  McDermid, McGreer, Melendez, Meneses-Goytia, Merloni, Merrifield, Meszaros,
  Meza, Minchev, Minniti, Mueller, Muller-Sanchez, Muna, Mu{\~{n}}oz, Myers,
  Nair, Nandra, Ness, Newman, Nichol, Nidever, Nitschelm, Noterdaeme,
  O’Connell, Oelkers, Oravetz, Oravetz, Ort{\'{i}}z, Osorio, Pace, Padilla,
  Palanque-Delabrouille, Palicio, Pan, Pan, Parikh, P{\^{a}}ris, Park, Peirani,
  Pellejero-Ibanez, Penny, Percival, Perez-Fournon, Petitjean, Pieri,
  Pinsonneault, Pisani, Prada, Prakash, de~Andrade~Queiroz, Raddick, Raichoor,
  Rembold, Richstein, Riffel, Riffel, Rix, Robin, Torres,
  Rom{\'{a}}n-Z{\'{u}}{\~{n}}iga, Ross, Rossi, Ruan, Ruggeri, Ruiz, Salvato,
  S{\'{a}}nchez, S{\'{a}}nchez, Almeida, S{\'{a}}nchez-Gallego, Rojas,
  Santiago, Schiavon, Schimoia, Schlafly, Schlegel, Schneider, Schuster,
  Schwope, Seo, Serenelli, Shen, Shen, Shetrone, Shull, Aguirre, Simon,
  Skrutskie, Slosar, Smethurst, Smith, Sobeck, Somers, Souter, Souto, Spindler,
  Stark, Stassun, Steinmetz, Stello, Storchi-Bergmann, Streblyanska,
  Stringfellow, Su{\'{a}}rez, Sun, Szigeti, Taghizadeh-Popp, Talbot, Tang, Tao,
  Tayar, Tembe, Teske, Thakar, Thomas, Tissera, Tojeiro, Tremonti, Troup, Urry,
  Valenzuela, Bosch, Vargas-Gonz{\'{a}}lez, Vargas-Maga{\~{n}}a, Vazquez,
  Villanova, Vogt, Wake, Wang, Weaver, Weijmans, Weinberg, Westfall, Whelan,
  Wilcots, Wild, Williams, Wilson, Wood-Vasey, Wylezalek, Xiao, Yan, Yang,
  Ybarra, Y{\`{e}}che, Zakamska, Zamora, Zarrouk, Zasowski, Zhang, Zhao, Zhao,
  Zheng, Zheng, Zhou, Zhu, Zinn, \& Zou}]{2018ApJS..235...42A}
Abolfathi, B., Aguado, D.~S., Aguilar, G., {et~al.} 2018, The Astrophysical
  Journal Supplement Series, 235, 42

\bibitem[{Allende~Prieto {et~al.}(2001)Allende~Prieto, Lambert, \&
  Asplund}]{2001ApJ...556L..63A}
Allende~Prieto, C., Lambert, D.~L., \& Asplund, M. 2001, The Astrophysical
  Journal, 556, L63

\bibitem[{Allende~Prieto {et~al.}(2002)Allende~Prieto, Lambert, \&
  Asplund}]{2002ApJ...573L.137A}
---. 2002, Astrophys. J., 573, L137

\bibitem[{Aller \& Czyzak(1983)}]{1983ApJS...51..211A}
Aller, L.~H., \& Czyzak, S.~J. 1983, The Astrophysical Journal Supplement
  Series, 51, 211

\bibitem[{{Astropy Collaboration} {et~al.}(2013){Astropy Collaboration},
  Robitaille, Tollerud, Greenfield, Droettboom, Bray, Aldcroft, Davis,
  Ginsburg, Price-Whelan, Kerzendorf, Conley, Crighton, Barbary, Muna,
  Ferguson, Grollier, Parikh, Nair, G{\"{u}}nther, Deil, Woillez, Conseil,
  Kramer, Turner, Singer, Fox, Weaver, Zabalza, Edwards, Azalee~Bostroem,
  Burke, Casey, Crawford, Dencheva, Ely, Jenness, Labrie, Lim, Pierfederici,
  Pontzen, Ptak, Refsdal, Servillat, \& Streicher}]{2013AA...558A..33A}
{Astropy Collaboration}, Robitaille, T.~P., Tollerud, E.~J., {et~al.} 2013,
  Astronomy {\&} Astrophysics, 558, A33

\bibitem[{Blackman(2004)}]{2004ASPC..313..401B}
Blackman, E.~G. 2004, in Asymmetrical Planetary Nebulae III: Winds, Structure
  and the Thunderbird, ed. M.~Meixner, J.~H. Kastner, B.~Balick, \& N.~Soker,
  Vol. 313, 401

\bibitem[{Bloecker(1995)}]{1995AA...299..755B}
Bloecker, T. 1995, Astronomy and Astrophysics, 299, 755

\bibitem[{Cardelli {et~al.}(1989)Cardelli, Clayton, \&
  Mathis}]{1989ApJ...345..245C}
Cardelli, J.~A., Clayton, G.~C., \& Mathis, J.~S. 1989, The Astrophysical
  Journal, 345, 245

\bibitem[{Cavichia {et~al.}(2017)Cavichia, Costa, Maciel, \&
  Molland{\'{a}}}]{2017MNRAS.468..272C}
Cavichia, O., Costa, R.~D., Maciel, W.~J., \& Molland{\'{a}}, M. 2017, Monthly
  Notices of the Royal Astronomical Society, 468, 272

\bibitem[{Ciardullo(2003)}]{2003astro.ph..1279C}
Ciardullo, R. 2003, Stellar Candles for the Extragalactic Distance Scale, 635,
  243

\bibitem[{Coccato \& {Coccato}(2016)}]{2016IAUFM..29B..20C}
Coccato, L., \& {Coccato}. 2016, Proceedings of the International Astronomical
  Union, 11, 20

\bibitem[{Corradi {et~al.}(2008)Corradi, Rodr{\'{i}}guez–Flores, Mampaso,
  Greimel, Viironen, Drew, Lennon, Mikolajewska, Sabin, \&
  Sokoloski}]{2008AA...480..409C}
Corradi, R. L.~M., Rodr{\'{i}}guez–Flores, E.~R., Mampaso, A., {et~al.} 2008,
  Astronomy {\&} Astrophysics, 480, 409

\bibitem[{Corradi {et~al.}(2010)Corradi, Valentini, Munari, Drew,
  Rodr{\'{i}}guez-Flores, Viironen, Greimel, Santander-Garc{\'{i}}a, Sabin,
  Mampaso, Parker, De~Pew, Sale, Unruh, Vink, Rodr{\'{i}}guez-Gil, Barlow,
  Lennon, Groot, Giammanco, Zijlstra, \& Walton}]{2010AA...509A..41C}
Corradi, R. L.~M., Valentini, M., Munari, U., {et~al.} 2010, Astronomy and
  Astrophysics, 509, A41

\bibitem[{Dalcanton {et~al.}(2015)Dalcanton, Fouesneau, Hogg, Lang, Leroy,
  Gordon, Sandstrom, Weisz, Williams, Bell, Dong, Gilbert, Gouliermis,
  Guhathakurta, Lauer, Schruba, Seth, \& Skillman}]{2015ApJ...814....3D}
Dalcanton, J.~J., Fouesneau, M., Hogg, D.~W., {et~al.} 2015, Astrophysical
  Journal, 814, 3

\bibitem[{De~Marco(2005)}]{2005AIPC..804..169D}
De~Marco, O. 2005, in AIP Conference Proceedings, ed. R.~Szczerba,
  G.~Stasi{\'{n}}ska, \& S.~K. Gorny, Vol. 804 (AIP), 169--172

\bibitem[{De~Marco(2009)}]{2009PASP..121..316D}
De~Marco, O. 2009, Publications of the Astronomical Society of the Pacific,
  121, 316

\bibitem[{Drew {et~al.}(2005)Drew, Greimel, Irwin, Aungwerojwit, Barlow,
  Corradi, Drake, Gansicke, Groot, Hales, Hopewell, Irwin, Knigge, Leisy,
  Lennon, Mampaso, Masheder, Matsuura, Morales-Rueda, Morris, Parker,
  Phillipps, Rodriguez-Gil, Roelofs, Skillen, Sokoloski, Steeghs, Unruh,
  Viironen, Vink, Walton, Witham, Wright, Zijlstra, \&
  Zurita}]{2005MNRAS.362..753D}
Drew, J.~E., Greimel, R., Irwin, M.~J., {et~al.} 2005, Monthly Notices of the
  Royal Astronomical Society, 362, 753

\bibitem[{Drew {et~al.}(2014)Drew, Gonzalez-solares, Greimel, Irwin,
  K{\"{u}}pc{\"{u}}~Yoldas, Lewis, Barentsen, Eisl{\"{o}}ffel, Farnhill,
  Martin, Walsh, Walton, Mohr-Smith, Raddi, Sale, Wright, Groot, Barlow,
  Corradi, Drake, Fabregat, Frew, G{\"{a}}nsicke, Knigge, Mampaso, Morris,
  Naylor, Parker, Phillipps, Ruhland, Steeghs, Unruh, Vink, Wesson, \&
  Zijlstra}]{2014MNRAS.440.2036D}
Drew, J.~E., Gonzalez-solares, E., Greimel, R., {et~al.} 2014, Monthly Notices
  of the Royal Astronomical Society, 440, 2036

\bibitem[{Fabian \& Hansen(1979)}]{1979MNRAS.187..283F}
Fabian, A.~C., \& Hansen, C.~J. 1979, Monthly Notices of the Royal Astronomical
  Society, 187, 283

\bibitem[{Ferland {et~al.}(2013)Ferland, Porter, van Hoof, Williams, Abel,
  Lykins, Shaw, Henney, \& Stancil}]{2013RMxAA..49..137F}
Ferland, G.~J., Porter, R.~L., van Hoof, P. A.~M., {et~al.} 2013, Revista
  Mexicana de Astronomia y Astrofisica, 49, 137

\bibitem[{Fragkou {et~al.}(2018)Fragkou, Parker, Bojicic, \&
  Aksaker}]{2018MNRAS.tmp.1882F}
Fragkou, V., Parker, Q.~A., Bojicic, I.~S., \& Aksaker, N. 2018, Monthly
  Notices of the Royal Astronomical Society, 480, 2916

\bibitem[{Frew(2008)}]{2008PhDT.......109F}
Frew, D.~J. 2008, PhD thesis, Department of Physics, Macquarie University, NSW
  2109, Australia, doi:10.13140/RG.2.1.2623.4406/1

\bibitem[{Frew {et~al.}(2014)Frew, Boji{\v{c}}i{\'{c}}, Parker, Pierce,
  Gunawardhana, \& Reid}]{2014MNRAS.440.1080F}
Frew, D.~J., Boji{\v{c}}i{\'{c}}, I.~S., Parker, Q.~A., {et~al.} 2014, Monthly
  Notices of the Royal Astronomical Society, 440, 1080

\bibitem[{Frew \& Parker(2010)}]{2010PASA...27..129F}
Frew, D.~J., \& Parker, Q.~A. 2010, Publications of the Astronomical Society of
  the Australia, 27, 129

\bibitem[{Garcia-Segura {et~al.}(1999)Garcia-Segura, Langer, Ro{\.{z}}yczka, \&
  Franco}]{1999ApJ...517..767G}
Garcia-Segura, G., Langer, N., Ro{\.{z}}yczka, M., \& Franco, J. 1999, The
  Astrophysical Journal, 517, 767

\bibitem[{Garcia‐Segura {et~al.}(2005)Garcia‐Segura, Lopez, \&
  Franco}]{2005ApJ...618..919G}
Garcia‐Segura, G., Lopez, J.~A., \& Franco, J. 2005, The Astrophysical
  Journal, 618, 919

\bibitem[{Gesicki {et~al.}(2018)Gesicki, Zijlstra, \&
  Miller~Bertolami}]{2018NatAs...2..580G}
Gesicki, K., Zijlstra, A.~A., \& Miller~Bertolami, M.~M. 2018, Nature
  Astronomy, 2, 580

\bibitem[{Gledhill {et~al.}(2018)Gledhill, Froebrich, Campbell-White, \&
  Jones}]{2018MNRAS.479.3759G}
Gledhill, T.~M., Froebrich, D., Campbell-White, J., \& Jones, A.~M. 2018,
  Monthly Notices of the Royal Astronomical Society, 479, 3759

\bibitem[{Graczyk {et~al.}(2014)Graczyk, Pietrzy{\'{n}}ski, Thompson, Gieren,
  Pilecki, Konorski, Udalski, Soszy{\'{n}}ski, Villanova, G{\'{o}}rski,
  Suchomska, Karczmarek, Kudritzki, Bresolin, \&
  Gallenne}]{2014ApJ...780...59G}
Graczyk, D., Pietrzy{\'{n}}ski, G., Thompson, I.~B., {et~al.} 2014,
  Astrophysical Journal, 780, 59

\bibitem[{Grevesse \& Sauval(1998)}]{1998SSRv...85..161G}
Grevesse, N., \& Sauval, A.~J. 1998, Space Science Reviews, 85, 161

\bibitem[{Gurzadian(1962)}]{1962VA......5...40G}
Gurzadian, G.~A. 1962, Vistas in Astronomy, 5, 40

\bibitem[{Holweger(2001)}]{2001AIPC..598...23H}
Holweger, H. 2001, in AIP Conference Proceedings, ed. R.~F.
  Wimmer-Schweingruber, Vol. 598, 23--30

\bibitem[{Hunter {et~al.}(2007)Hunter, Asnicar, Manghi, Pasolli, Tett, \&
  Segata}]{Hunter:2007}
Hunter, J.~D., Asnicar, F., Manghi, P., {et~al.} 2007, Computing in Science
  {\&} Engineering, 9, 90

\bibitem[{Hurley-Keller \& Morrison(2004)}]{2004ApJ...616..804H}
Hurley-Keller, D., \& Morrison, H.~L. 2004, The Astrophysical Journal, 616, 804

\bibitem[{Iben(1995)}]{1995PhR...250....1I}
Iben, I. 1995, Physics Reports, 250, 1

\bibitem[{I{\l}kiewicz \& Miko{\l}ajewska(2017)}]{2017AA...606A.110I}
I{\l}kiewicz, K., \& Miko{\l}ajewska, J. 2017, Astronomy {\&} Astrophysics,
  606, A110

\bibitem[{Jacob {et~al.}(2013)Jacob, Sch{\"{o}}nberner, \&
  Steffen}]{2013AA...558A..78J}
Jacob, R., Sch{\"{o}}nberner, D., \& Steffen, M. 2013, Astronomy {\&}
  Astrophysics, 558, A78

\bibitem[{Jacoby \& Ford(1986)}]{1986ApJ...304..490J}
Jacoby, G.~H., \& Ford, H.~C. 1986, The Astrophysical Journal, 304, 490

\bibitem[{Jones {et~al.}(2001)Jones, Oliphant, Peterson, \&
  {Others}}]{jones_scipy_2001}
Jones, E., Oliphant, T., Peterson, P., \& {Others}. 2001, {SciPy: Open source
  scientific tools for Python}

\bibitem[{Khromov(1989)}]{1989SSRv...51..339K}
Khromov, G.~S. 1989, Space Science Reviews, 51, 339

\bibitem[{Kniazev {et~al.}(2014)Kniazev, Grebel, Zucker, Rix,
  Mart{\'{i}}nez-Delgado, \& Snedden}]{2014AJ....147...16K}
Kniazev, A.~Y., Grebel, E.~K., Zucker, D.~B., {et~al.} 2014, Astronomical
  Journal, 147, 16

\bibitem[{Kniazev {et~al.}(2008)Kniazev, Zijlstra, Grebel, Pilyugin, Pustilnik,
  V{\"{a}}is{\"{a}}nen, Buckley, Hashimoto, Loaring, Romero, Still, Burgh, \&
  Nordsieck}]{2008MNRAS.388.1667K}
Kniazev, A.~Y., Zijlstra, A.~A., Grebel, E.~K., {et~al.} 2008, Monthly Notices
  of the Royal Astronomical Society, 388, 1667

\bibitem[{Kwitter {et~al.}(2012)Kwitter, Lehman, Balick, \&
  Henry}]{2012ApJ...753...12K}
Kwitter, K.~B., Lehman, E.~M., Balick, B., \& Henry, R.~B. 2012, Astrophysical
  Journal, 753, 12

\bibitem[{Kwok {et~al.}(1978)Kwok, Purton, \& Fitzgerald}]{1978ApJ...219L.125K}
Kwok, S., Purton, C.~R., \& Fitzgerald, P.~M. 1978, The Astrophysical Journal,
  219, L125

\bibitem[{{LSST Science Collaboration} {et~al.}(2009){LSST Science
  Collaboration}, Abell, Allison, Anderson, Andrew, Angel, Armus, Arnett,
  Asztalos, Axelrod, Bailey, Ballantyne, Bankert, Barkhouse, Barr, Barrientos,
  Barth, Bartlett, Becker, Becla, Beers, Bernstein, Biswas, Blanton, Bloom,
  Bochanski, Boeshaar, Borne, Bradac, Brandt, Bridge, Brown, Brunner, Bullock,
  Burgasser, Burge, Burke, Cargile, Chandrasekharan, Chartas, Chesley, Chu,
  Cinabro, Claire, Claver, Clowe, Connolly, Cook, Cooke, Cooray, Covey,
  Culliton, de~Jong, de~Vries, Debattista, Delgado, Dell'Antonio, Dhital,
  Di~Stefano, Dickinson, Dilday, Djorgovski, Dobler, Donalek, Dubois-Felsmann,
  Durech, Eliasdottir, Eracleous, Eyer, Falco, Fan, Fassnacht, Ferguson,
  Fernandez, Fields, Finkbeiner, Figueroa, Fox, Francke, Frank, Frieman,
  Fromenteau, Furqan, Galaz, Gal-Yam, Garnavich, Gawiser, Geary, Gee, Gibson,
  Gilmore, Grace, Green, Gressler, Grillmair, Habib, Haggerty, Hamuy, Harris,
  Hawley, Heavens, Hebb, Henry, Hileman, Hilton, Hoadley, Holberg, Holman,
  Howell, Infante, Ivezic, Jacoby, Jain, {R}, {Jedicke}, Jee, Jernigan, Jha,
  Johnston, Jones, Juric, Kaasalainen, {Styliani}, {Kafka}, Kahn, Kaib,
  Kalirai, Kantor, Kasliwal, Keeton, Kessler, Knezevic, Kowalski, Krabbendam,
  Krughoff, Kulkarni, Kuhlman, Lacy, Lepine, Liang, Lien, Lira, Long, Lorenz,
  Lotz, Lupton, Lutz, Macri, Mahabal, Mandelbaum, Marshall, May, McGehee,
  Meadows, Meert, Milani, Miller, Miller, Mills, Minniti, Monet, Mukadam,
  Nakar, Neill, Newman, Nikolaev, Nordby, O'Connor, Oguri, Oliver, Olivier,
  Olsen, Olsen, Olszewski, Oluseyi, Padilla, Parker, Pepper, Peterson, Petry,
  Pinto, Pizagno, Popescu, Prsa, Radcka, Raddick, Rasmussen, Rau, Rho, Rhoads,
  Richards, Ridgway, Robertson, Roskar, Saha, Sarajedini, Scannapieco, Schalk,
  Schindler, Schmidt, Schmidt, Schneider, Schumacher, Scranton, Sebag, Seppala,
  Shemmer, Simon, Sivertz, Smith, Smith, Smith, Spitz, Stanford, Stassun,
  Strader, Strauss, Stubbs, Sweeney, Szalay, Szkody, Takada, Thorman, Trilling,
  Trimble, Tyson, Van~Berg, Berk, VanderPlas, Verde, Vrsnak, Walkowicz,
  Wandelt, Wang, Wang, Warner, Wechsler, West, Wiecha, Williams, Willman,
  Wittman, Wolff, Wood-Vasey, Wozniak, Young, Zentner, \&
  Zhan}]{2009arXiv0912.0201L}
{LSST Science Collaboration}, Abell, P.~A., Allison, J., {et~al.} 2009, LSST
  Corporation

\bibitem[{Magrini \& Gon{\c{c}}alves(2009)}]{2009MNRAS.398..280M}
Magrini, L., \& Gon{\c{c}}alves, D.~R. 2009, Monthly Notices of the Royal
  Astronomical Society, 398, 280

\bibitem[{Majaess {et~al.}(2018)Majaess, D{\'{e}}k{\'{a}}ny, Hajdu, Minniti,
  Turner, \& Gieren}]{2018ApSS.363..127M}
Majaess, D., D{\'{e}}k{\'{a}}ny, I., Hajdu, G., {et~al.} 2018, Astrophysics and
  Space Science, 363, 127

\bibitem[{Matt {et~al.}(2004)Matt, Ls, Frank, \&
  Blackman}]{2004ASPC..313..449M}
Matt, S., Ls, C., Frank, A., \& Blackman, E. 2004, in Asymmetrical Planetary
  Nebulae III: Winds, Structure and the Thunderbird, ed. M.~Meixner, J.~H.
  Kastner, B.~Balick, \& N.~Soker, Vol. 313, 449--455

\bibitem[{Merrett {et~al.}(2006)Merrett, Merrifield, Douglas, Kuijken,
  Romanowsky, Napolitano, Arnaboldi, Capaccioli, Freeman, Gerhard, Coccato,
  Carter, Evans, Wilkinson, Halliday, \& Bridges}]{2006MNRAS.369..120M}
Merrett, H.~R., Merrifield, M.~R., Douglas, N.~G., {et~al.} 2006, Monthly
  Notices of the Royal Astronomical Society, 369, 120

\bibitem[{Miller~Bertolami(2016)}]{2016AA...588A..25M}
Miller~Bertolami, M.~M. 2016, Astronomy {\&} Astrophysics, 588, A25

\bibitem[{Moe \& De~Marco(2006)}]{2006ApJ...650..916M}
Moe, M., \& De~Marco, O. 2006, The Astrophysical Journal, 650, 916

\bibitem[{Morisset(2013)}]{2013ascl.soft04020M}
Morisset, C. 2013, {pyCloudy: Tools to manage astronomical Cloudy
  photoionization}, Astrophysics Source Code Library

\bibitem[{Morris \& Montez(2015)}]{2015AAS...22513907M}
Morris, M., \& Montez, R. 2015, in American Astronomical Society Meeting
  Abstracts, Vol. 225, American Astronomical Society Meeting Abstracts {\#}225,
  139.07

\bibitem[{Nolthenius \& Ford(1987)}]{1987ApJ...317...62N}
Nolthenius, R., \& Ford, H.~C. 1987, The Astrophysical Journal, 317, 62

\bibitem[{Nordhaus {et~al.}(2007)Nordhaus, Blackman, \&
  Frank}]{2007MNRAS.376..599N}
Nordhaus, J., Blackman, E.~G., \& Frank, A. 2007, Monthly Notices of the Royal
  Astronomical Society, 376, 599

\bibitem[{Oke \& Gunn(1983)}]{1983ApJ...266..713O}
Oke, J.~B., \& Gunn, J.~E. 1983, The Astrophysical Journal, 266, 713

\bibitem[{Parker {et~al.}(2016)Parker, Boji{\v{c}}i{\'{c}}, \&
  Frew}]{2016JPhCS.728c2008P}
Parker, Q.~A., Boji{\v{c}}i{\'{c}}, I.~S., \& Frew, D.~J. 2016, in Journal of
  Physics Conference Series, Vol. 728, Journal of Physics Conference Series,
  32008

\bibitem[{Parker {et~al.}(2005)Parker, Phillipps, Pierce, Hartley, Hambly,
  Read, MacGillivray, Tritton, Cass, Cannon, Cohen, Drew, Frew, Hopewell,
  Mader, Malin, Masheder, Morgan, Morris, Russeil, Russell, \&
  Walker}]{2005MNRAS.362..689P}
Parker, Q.~A., Phillipps, S., Pierce, M.~J., {et~al.} 2005, Monthly Notices of
  the Royal Astronomical Society, 362, 689

\bibitem[{Pastorello {et~al.}(2013)Pastorello, Sarzi, Cappellari, Emsellem,
  Mamon, Bacon, Davies, \& Tim~de Zeeuw}]{2013MNRAS.430.1219P}
Pastorello, N., Sarzi, M., Cappellari, M., {et~al.} 2013, Monthly Notices of
  the Royal Astronomical Society, 430, 1219

\bibitem[{Peng {et~al.}(2004)Peng, Ford, \& Freeman}]{2004ApJ...602..705P}
Peng, E.~W., Ford, H.~C., \& Freeman, K.~C. 2004, The Astrophysical Journal,
  602, 705

\bibitem[{P{\'{e}}rez \& Granger(2007)}]{2007CSE.....9c..21P}
P{\'{e}}rez, F., \& Granger, B.~E. 2007, Computing in Science and Engineering,
  9, 21

\bibitem[{Perinotto {et~al.}(2004)Perinotto, Sch{\"{o}}nberner, Steffen, \&
  Calonaci}]{2004AA...414..993P}
Perinotto, M., Sch{\"{o}}nberner, D., Steffen, M., \& Calonaci, C. 2004,
  Astronomy {\&} Astrophysics, 414, 993

\bibitem[{Phillips(2007)}]{2007MNRAS.378..231P}
Phillips, J.~P. 2007, Monthly Notices of the Royal Astronomical Society, 378,
  231

\bibitem[{Pietrzy{\'{n}}ski {et~al.}(2013)Pietrzy{\'{n}}ski, Graczyk, Gieren,
  Thompson, Pilecki, Udalski, Soszy{\'{n}}ski, Koz{\l}owski, Konorski,
  Suchomska, Bono, Moroni, Villanova, Nardetto, Bresolin, Kudritzki, Storm,
  Gallenne, Smolec, Minniti, Kubiak, Szyma{\'{n}}ski, Poleski, {Wyrzykowski},
  Ulaczyk, Pietrukowicz, G{\'{o}}rski, \& Karczmarek}]{2013Natur.495...76P}
Pietrzy{\'{n}}ski, G., Graczyk, D., Gieren, W., {et~al.} 2013, Nature, 495, 76

\bibitem[{Richer {et~al.}(1998)Richer, McCall, \&
  Stasi{\'{n}}ska}]{1998AA...340...67R}
Richer, M., McCall, M.~L., \& Stasi{\'{n}}ska, G. 1998, Astronomy and
  Astrophysics, 340, 67

\bibitem[{Rodr{\'{i}}guez-Flores {et~al.}(2014)Rodr{\'{i}}guez-Flores, Corradi,
  Mampaso, Garc{\'{i}}a-Alvarez, Munari, Greimel, Rubio-D{\'{i}}ez, \&
  Santander-Garc{\'{i}}a}]{2014AA...567A..49R}
Rodr{\'{i}}guez-Flores, E.~R., Corradi, R. L.~M., Mampaso, A., {et~al.} 2014,
  Astronomy and Astrophysics, 567, A49

\bibitem[{Sanders {et~al.}(2012)Sanders, Caldwell, McDowell, \&
  Harding}]{2012ApJ...758..133S}
Sanders, N.~E., Caldwell, N., McDowell, J., \& Harding, P. 2012, Astrophysical
  Journal, 758, 133

\bibitem[{Saviane {et~al.}(2009)Saviane, Exter, Tsamis, Gallart, \&
  P{\'{e}}quignot}]{2009AA...494..515S}
Saviane, I., Exter, K., Tsamis, Y., Gallart, C., \& P{\'{e}}quignot, D. 2009,
  Astronomy {\&} Astrophysics, 494, 515

\bibitem[{Schmeja \& Kimeswenger(2001)}]{2001AA...377L..18S}
Schmeja, S., \& Kimeswenger, S. 2001, Astronomy {\&} Astrophysics, 377, L18

\bibitem[{Sch{\"{o}}nberner {et~al.}(2005)Sch{\"{o}}nberner, Jacob, \&
  Steffen}]{2005AA...441..573S}
Sch{\"{o}}nberner, D., Jacob, R., \& Steffen, M. 2005, Astronomy {\&}
  Astrophysics, 441, 573

\bibitem[{Soker(1997)}]{1997ApJS..112..487S}
Soker, N. 1997, The Astrophysical Journal Supplement Series, 112, 487

\bibitem[{Soker(2006)}]{2006PASP..118..260S}
---. 2006, Publications of the Astronomical Society of the Pacific, 118, 260

\bibitem[{Su{\'{a}}rez {et~al.}(2006)Su{\'{a}}rez, Garc{\'{i}}a-Lario,
  Manchado, Manteiga, Ulla, \& Pottasch}]{2006AA...458..173S}
Su{\'{a}}rez, O., Garc{\'{i}}a-Lario, P., Manchado, A., {et~al.} 2006,
  Astronomy {\&} Astrophysics, 458, 173

\bibitem[{Toal{\'{a}} \& Arthur(2014)}]{2014MNRAS.443.3486T}
Toal{\'{a}}, J.~A., \& Arthur, S.~J. 2014, Monthly Notices of the Royal
  Astronomical Society, 443, 3486

\bibitem[{van~den Bergh(1999)}]{1999AARv...9..273V}
van~den Bergh, S. 1999, Astronomy and Astrophysics Review, 9, 273

\bibitem[{Van Der~Walt {et~al.}(2011)Van Der~Walt, Colbert, \&
  Varoquaux}]{van2011numpy}
Van Der~Walt, S., Colbert, S.~C., \& Varoquaux, G. 2011, Computing in Science
  and Engineering, doi:10.1109/MCSE.2011.37

\bibitem[{Vassiliadis \& Wood(1994)}]{1994ApJS...92..125V}
Vassiliadis, E., \& Wood, P.~R. 1994, The Astrophysical Journal Supplement
  Series, 92, 125

\bibitem[{Veyette {et~al.}(2014)Veyette, Williams, Dalcanton, Balick, Caldwell,
  Fouesneau, Girardi, Gordon, Kalirai, Rosenfield, \&
  Seth}]{2014ApJ...792..121V}
Veyette, M.~J., Williams, B.~F., Dalcanton, J.~J., {et~al.} 2014, The
  Astrophysical Journal, 792, 121

\bibitem[{Walsh {et~al.}(1997)Walsh, Dudziak, Minniti, \&
  Zijlstra}]{1997ApJ...487..651W}
Walsh, J.~R., Dudziak, G., Minniti, D., \& Zijlstra, A.~A. 1997, The
  Astrophysical Journal, 487, 651

\bibitem[{Weston {et~al.}(2009)Weston, Napiwotzki, \&
  Sale}]{2009JPhCS.172a2033W}
Weston, S., Napiwotzki, R., \& Sale, S. 2009, Journal of Physics: Conference
  Series, 172, 012033

\bibitem[{Yuan \& Liu(2013)}]{2013MNRAS.436..718Y}
Yuan, H.~B., \& Liu, X.~W. 2013, Monthly Notices of the Royal Astronomical
  Society, 436, 718

\end{thebibliography}

\setlength\mylength{\dimexpr\textwidth-5\arrayrulewidth-8\tabcolsep}
\begin{longtable*}{C{.07\mylength}C{.07\mylength}C{.07\mylength}C{.07\mylength}C{.07\mylength}C{.07\mylength}C{.07\mylength}}
\multicolumn{2}{c}{\text{Variant I:}} & \multicolumn{3}{c}{\text{$\Omega_{\rm aperture}/\Omega_{\rm neb} = 1.3$}} &\multicolumn{2}{c}{\text{$f_{neb}$ = $1$}} \\
\hline
Age & u & g & r & i & z & y\\
(yrs)  &   (mags) &   (mags) &   (mags) &   (mags) &   (mags) &   (mags) \\
\hline
3502 & -2.59 & -3.63 & -3.38 & -2.20 & -2.18 & -2.09 \\
4154 & -2.70 & -3.91 & -3.47 & -2.21 & -2.20 & -2.15 \\
4860 & -2.46 & -3.94 & -2.98 & -1.87 & -1.98 & -1.94 \\
5663 & -2.11 & -3.66 & -2.53 & -1.44 & -1.62 & -1.66 \\
6328 & -1.93 & -3.41 & -2.36 & -1.21 & -1.40 & -1.49 \\
6745 & -1.95 & -3.23 & -2.43 & -1.14 & -1.29 & -1.39 \\
7031 & -1.98 & -3.00 & -2.53 & -1.06 & -1.13 & -1.28 \\
7224 & -2.00 & -2.72 & -2.64 & -0.98 & -0.98 & -1.21 \\
7387 & -1.71 & -2.10 & -2.49 & -0.63 & -0.54 & -0.95 \\
7453 & -1.44 & -1.60 & -2.27 & -0.33 & -0.18 & -0.72 \\
7552 & -1.09 & -0.90 & -1.98 &  0.08 &  0.28 & -0.45 \\
7751 & -0.75 & -0.22 & -1.68 &  0.45 &  0.71 & -0.24 \\
8351 & -0.51 &  0.24 & -1.46 &  0.73 &  1.00 & -0.11 \\
\hline
\multicolumn{2}{c}{\text{Variant II:}} & \multicolumn{3}{c}{\text{$\Omega_{\rm aperture}/\Omega_{\rm neb} = 0.9$}} & \multicolumn{2}{c}{\text{$f_{neb}$ = $4\times10^{-1}$}} \\
\hline
3502 & -1.88 & -2.78 & -2.52 & -1.39 & -1.35 & -1.24 \\
4154 & -1.90 & -3.03 & -2.58 & -1.35 & -1.33 & -1.27 \\
4860 & -1.62 & -3.04 & -2.09 & -0.99 & -1.10 & -1.05 \\
5663 & -1.25 & -2.76 & -1.63 & -0.56 & -0.73 & -0.77 \\
6328 & -1.06 & -2.51 & -1.46 & -0.32 & -0.50 & -0.59 \\
6745 & -1.06 & -2.33 & -1.54 & -0.24 & -0.39 & -0.49 \\
7031 & -1.09 & -2.10 & -1.63 & -0.16 & -0.24 & -0.38 \\
7224 & -1.10 & -1.82 & -1.74 & -0.08 & -0.08 & -0.31 \\
7387 & -0.82 & -1.20 & -1.59 &  0.27 &  0.36 & -0.05 \\
7453 & -0.55 & -0.70 & -1.37 &  0.57 &  0.72 &  0.19 \\
7552 & -0.20 &  0.00 & -1.07 &  0.97 &  1.18 &  0.45 \\
7751 &  0.14 &  0.67 & -0.78 &  1.35 &  1.60 &  0.66 \\
8351 &  0.38 &  1.13 & -0.56 &  1.62 &  1.89 &  0.79 \\
\hline
\multicolumn{2}{c}{\text{Variant III:}} & \multicolumn{3}{c}{\text{$\Omega_{\rm aperture}/\Omega_{\rm neb} = 0.7$}} & \multicolumn{2}{c}{\text{$f_{neb}$ = $2\times10^{-1}$}} \\
\hline
3502 & -1.45 & -2.17 & -1.90 & -0.83 & -0.77 & -0.65 \\
4154 & -1.35 & -2.38 & -1.93 & -0.74 & -0.70 & -0.63 \\
4860 & -1.03 & -2.38 & -1.43 & -0.36 & -0.45 & -0.39 \\
5663 & -0.63 & -2.09 & -0.97 &  0.10 & -0.07 & -0.10 \\
6328 & -0.41 & -1.84 & -0.79 &  0.34 &  0.17 &  0.08 \\
6745 & -0.41 & -1.66 & -0.86 &  0.43 &  0.29 &  0.18 \\
7031 & -0.42 & -1.42 & -0.95 &  0.51 &  0.44 &  0.30 \\
7224 & -0.44 & -1.14 & -1.06 &  0.59 &  0.60 &  0.37 \\
7387 & -0.15 & -0.53 & -0.91 &  0.94 &  1.04 &  0.63 \\
7453 &  0.12 & -0.02 & -0.69 &  1.24 &  1.39 &  0.86 \\
7552 &  0.47 &  0.67 & -0.39 &  1.64 &  1.85 &  1.13 \\
7751 &  0.81 &  1.34 & -0.10 &  2.02 &  2.27 &  1.34 \\
8351 &  1.04 &  1.79 &  0.12 &  2.29 &  2.56 &  1.47 \\
\hline
\multicolumn{2}{c}{\text{Variant IV:}} & \multicolumn{3}{c}{\text{$\Omega_{\rm aperture}/\Omega_{\rm neb} = 0.5$}} & \multicolumn{2}{c}{\text{$f_{neb}$ = $1\times10^{-1}$}} \\
\hline
3502 & -1.08 & -1.53 & -1.24 & -0.27 & -0.17 & -0.03 \\
4154 & -0.82 & -1.65 & -1.20 & -0.08 & -0.02 &  0.07 \\
4860 & -0.42 & -1.61 & -0.68 &  0.35 &  0.28 &  0.35 \\
5663 &  0.05 & -1.31 & -0.20 &  0.83 &  0.69 &  0.67 \\
6328 &  0.31 & -1.05 & -0.01 &  1.10 &  0.94 &  0.87 \\
6745 &  0.34 & -0.86 & -0.07 &  1.20 &  1.07 &  0.97 \\
7031 &  0.34 & -0.62 & -0.16 &  1.29 &  1.22 &  1.09 \\
7224 &  0.34 & -0.34 & -0.26 &  1.38 &  1.39 &  1.17 \\
7387 &  0.63 &  0.27 & -0.11 &  1.72 &  1.82 &  1.42 \\
7453 &  0.90 &  0.76 &  0.11 &  2.03 &  2.18 &  1.66 \\
7552 &  1.24 &  1.46 &  0.41 &  2.42 &  2.64 &  1.93 \\
7751 &  1.58 &  2.11 &  0.70 &  2.80 &  3.05 &  2.14 \\
8351 &  1.81 &  2.55 &  0.92 &  3.07 &  3.34 &  2.27 \\
\hline
\multicolumn{2}{c}{\text{Variant V:}} & \multicolumn{3}{c}{\text{$\Omega_{\rm aperture}/\Omega_{\rm neb} = 0.3$}} & \multicolumn{2}{c}{\text{$f_{neb}$ = $4\times10^{-2}$}} \\
\hline
3502 & -0.77 & -0.83 & -0.47 &  0.27 &  0.45 &  0.63 \\
4154 & -0.31 & -0.73 & -0.27 &  0.67 &  0.80 &  0.93 \\
4860 &  0.23 & -0.58 &  0.29 &  1.20 &  1.21 &  1.31 \\
5663 &  0.82 & -0.23 &  0.82 &  1.77 &  1.69 &  1.71 \\
6328 &  1.19 &  0.05 &  1.06 &  2.11 &  1.99 &  1.94 \\
6745 &  1.30 &  0.25 &  1.03 &  2.24 &  2.15 &  2.07 \\
7031 &  1.35 &  0.49 &  0.96 &  2.36 &  2.31 &  2.20 \\
7224 &  1.38 &  0.77 &  0.86 &  2.45 &  2.48 &  2.28 \\
7387 &  1.68 &  1.37 &  1.02 &  2.80 &  2.92 &  2.54 \\
7453 &  1.94 &  1.86 &  1.24 &  3.10 &  3.26 &  2.78 \\
7552 &  2.29 &  2.52 &  1.54 &  3.49 &  3.71 &  3.05 \\
7751 &  2.61 &  3.14 &  1.83 &  3.85 &  4.11 &  3.25 \\
8351 &  2.83 &  3.54 &  2.05 &  4.11 &  4.38 &  3.39 \\
\hline
\multicolumn{2}{c}{\text{Variant VI:}} & \multicolumn{3}{c}{\text{$\Omega_{\rm aperture}/\Omega_{\rm neb} = 0.15$}} & \multicolumn{2}{c}{\text{$f_{neb}$ = $9\times10^{-3}$}} \\
\hline
3502 & -0.62 & -0.32 &  0.10 &  0.61 &  0.85 &  1.08 \\
4154 & -0.01 &  0.11 &  0.59 &  1.23 &  1.45 &  1.65 \\
4860 &  0.66 &  0.51 &  1.24 &  1.90 &  2.06 &  2.23 \\
5663 &  1.41 &  1.01 &  1.91 &  2.63 &  2.72 &  2.82 \\
6328 &  1.96 &  1.38 &  2.29 &  3.13 &  3.16 &  3.20 \\
6745 &  2.23 &  1.62 &  2.36 &  3.38 &  3.40 &  3.39 \\
7031 &  2.40 &  1.88 &  2.35 &  3.56 &  3.60 &  3.56 \\
7224 &  2.50 &  2.15 &  2.30 &  3.70 &  3.79 &  3.67 \\
7387 &  2.84 &  2.73 &  2.48 &  4.06 &  4.22 &  3.95 \\
7453 &  3.10 &  3.17 &  2.70 &  4.35 &  4.54 &  4.18 \\
7552 &  3.42 &  3.75 &  3.00 &  4.71 &  4.95 &  4.45 \\
7751 &  3.71 &  4.25 &  3.28 &  5.03 &  5.30 &  4.66 \\
8351 &  3.90 &  4.55 &  3.50 &  5.25 &  5.54 &  4.80 \\
\hline
\multicolumn{2}{c}{\text{Variant VII:}} & \multicolumn{3}{c}{\text{$\Omega_{\rm aperture}/\Omega_{\rm neb} = 0.1$}} & \multicolumn{2}{c}{\text{$f_{neb}$ = $4\times10^{-3}$}} \\
\hline
3502 & -0.59 & -0.21 &  0.24 &  0.68 &  0.94 &  1.18 \\
4154 &  0.06 &  0.35 &  0.83 &  1.36 &  1.61 &  1.84 \\
4860 &  0.76 &  0.89 &  1.54 &  2.09 &  2.30 &  2.52 \\
5663 &  1.55 &  1.50 &  2.28 &  2.88 &  3.05 &  3.22 \\
6328 &  2.17 &  1.96 &  2.77 &  3.47 &  3.60 &  3.70 \\
6745 &  2.52 &  2.24 &  2.94 &  3.80 &  3.90 &  3.96 \\
7031 &  2.75 &  2.51 &  3.00 &  4.03 &  4.14 &  4.16 \\
7224 &  2.91 &  2.79 &  2.99 &  4.20 &  4.35 &  4.30 \\
7387 &  3.27 &  3.33 &  3.20 &  4.57 &  4.77 &  4.60 \\
7453 &  3.52 &  3.73 &  3.42 &  4.85 &  5.07 &  4.84 \\
7552 &  3.83 &  4.22 &  3.71 &  5.18 &  5.44 &  5.11 \\
7751 &  4.09 &  4.64 &  3.99 &  5.47 &  5.76 &  5.32 \\
8351 &  4.27 &  4.88 &  4.20 &  5.67 &  5.96 &  5.45 \\
\hline
\multicolumn{2}{c}{\text{Variant VIII:}} & \multicolumn{3}{c}{\text{$\Omega_{\rm aperture}/\Omega_{\rm neb} = 0.075$}} & \multicolumn{2}{c}{\text{$f_{neb}$ = $2\times10^{-3}$}} \\
\hline
3502 & -0.58 & -0.16 &  0.30 &  0.71 &  0.98 &  1.23 \\
4154 &  0.08 &  0.47 &  0.96 &  1.42 &  1.69 &  1.93 \\
4860 &  0.80 &  1.10 &  1.70 &  2.18 &  2.43 &  2.67 \\
5663 &  1.62 &  1.81 &  2.50 &  3.01 &  3.24 &  3.45 \\
6328 &  2.28 &  2.35 &  3.08 &  3.66 &  3.86 &  4.02 \\
6745 &  2.67 &  2.68 &  3.34 &  4.04 &  4.21 &  4.35 \\
7031 &  2.95 &  2.97 &  3.47 &  4.31 &  4.49 &  4.59 \\
7224 &  3.15 &  3.24 &  3.52 &  4.52 &  4.72 &  4.76 \\
7387 &  3.52 &  3.74 &  3.76 &  4.90 &  5.13 &  5.09 \\
7453 &  3.77 &  4.10 &  3.99 &  5.17 &  5.42 &  5.32 \\
7552 &  4.07 &  4.52 &  4.28 &  5.48 &  5.75 &  5.59 \\
7751 &  4.32 &  4.86 &  4.55 &  5.74 &  6.03 &  5.80 \\
8351 &  4.48 &  5.07 &  4.74 &  5.92 &  6.21 &  5.94 \\
\hline
\multicolumn{2}{c}{\text{Variant IX:}} & \multicolumn{3}{c}{\text{$\Omega_{\rm aperture}/\Omega_{\rm neb} = 0.05$}} & \multicolumn{2}{c}{\text{$f_{neb}$ = $1\times10^{-3}$}} \\
\hline
3502 & -0.57 & -0.13 &  0.34 &  0.72 &  1.00 &  1.25 \\
4154 &  0.10 &  0.53 &  1.02 &  1.45 &  1.73 &  1.98 \\
4860 &  0.82 &  1.22 &  1.78 &  2.22 &  2.49 &  2.74 \\
5663 &  1.65 &  1.98 &  2.61 &  3.08 &  3.34 &  3.57 \\
6328 &  2.33 &  2.58 &  3.24 &  3.76 &  4.00 &  4.20 \\
6745 &  2.75 &  2.95 &  3.57 &  4.17 &  4.39 &  4.57 \\
7031 &  3.05 &  3.25 &  3.77 &  4.47 &  4.69 &  4.85 \\
7224 &  3.28 &  3.53 &  3.88 &  4.70 &  4.93 &  5.05 \\
7387 &  3.66 &  4.00 &  4.16 &  5.09 &  5.34 &  5.40 \\
7453 &  3.91 &  4.32 &  4.39 &  5.34 &  5.61 &  5.64 \\
7552 &  4.19 &  4.69 &  4.68 &  5.64 &  5.92 &  5.90 \\
7751 &  4.43 &  4.98 &  4.93 &  5.89 &  6.18 &  6.12 \\
8351 &  4.59 &  5.16 &  5.11 &  6.05 &  6.35 &  6.26 \\
\hline
\multicolumn{2}{c}{\text{Variant X:}} & \multicolumn{3}{c}{\text{$\Omega_{\rm aperture}/\Omega_{\rm neb} = 0.01$}} & \multicolumn{2}{c}{\text{$f_{neb}$ = $3\times10^{-5}$}} \\
\hline
3502 & -0.57 & -0.10 &  0.37 &  0.74 &  1.02 &  1.28 \\
4154 &  0.11 &  0.60 &  1.10 &  1.48 &  1.77 &  2.03 \\
4860 &  0.85 &  1.36 &  1.88 &  2.27 &  2.57 &  2.83 \\
5663 &  1.69 &  2.22 &  2.75 &  3.16 &  3.45 &  3.72 \\
6328 &  2.39 &  2.93 &  3.47 &  3.88 &  4.18 &  4.45 \\
6745 &  2.84 &  3.38 &  3.92 &  4.34 &  4.64 &  4.91 \\
7031 &  3.17 &  3.72 &  4.26 &  4.68 &  4.98 &  5.25 \\
7224 &  3.44 &  3.98 &  4.52 &  4.95 &  5.25 &  5.52 \\
7387 &  3.84 &  4.39 &  4.91 &  5.34 &  5.65 &  5.91 \\
7453 &  4.08 &  4.63 &  5.15 &  5.59 &  5.89 &  6.15 \\
7552 &  4.36 &  4.91 &  5.43 &  5.86 &  6.16 &  6.42 \\
7751 &  4.58 &  5.13 &  5.65 &  6.08 &  6.38 &  6.64 \\
8351 &  4.72 &  5.27 &  5.79 &  6.22 &  6.52 &  6.78 \\
\hline
\caption{Absolute magnitudes for all 13 PN models for corresponding $\Omega_{\rm aperture}/\Omega_{\rm neb}$ and $f_{neb}$}
\label{table: neb_fraction}
\end{longtable*}

\end{document}